\documentclass{aa}
\usepackage[varg]{txfonts}
\usepackage{graphicx}	
\usepackage{amsmath}	
\usepackage{amssymb}	

\begin{document}

\title{Exoplanet X-ray irradiation and evaporation rates with eROSITA}

\author{G. Foster\inst{1, 2}
   \and K. Poppenhaeger\inst{1,2}
     \and N. Ilic\inst{1, 2}
     \and A. Schwope\inst{1}}

\institute{Leibniz Institute for Astrophysics Potsdam, An der Sternwarte 16, 14482 Potsdam, Germany
  \and Potsdam University, Institute for Physics and Astronomy, Karl-Liebknecht-Straße 24/25, 14476 Potsdam, Germany}

\date{Accepted June 21 2021, submitted April 15 2021}

\abstract{High-energy irradiation is a driver for atmospheric evaporation and mass loss in exoplanets. This work is based on data from eROSITA, the soft X-ray instrument aboard SRG 
(Spectrum Roentgen Gamma) mission, 
as well as archival data from other missions, we aim to characterise the high-energy environment of known exoplanets and estimate their mass loss rates. We use X-ray source catalogues from eROSITA, \textit{XMM-Newton}, \textit{Chandra} and ROSAT to derive X-ray luminosities of exoplanet host stars in the 0.2-2 keV energy band with an underlying coronal, i.e.\ optically thin thermal spectrum. We present a catalogue of stellar X-ray and EUV luminosities, exoplanetary X-ray and EUV irradiation fluxes and estimated mass loss rates for a 
total of 287 
exoplanets, 96 
among them being characterised for the first time from new eROSITA detections. We identify 14 first time X-ray detections of 
transiting exoplanets that are 
subject to irradiation levels known to cause observable evaporation signatures in other exoplanets, which makes them suitable targets for follow-up observations.
}


\maketitle

\section{Introduction}

Exoplanets have been detected in a wide variety of orbital architectures, and a significant fraction of them orbit their host stars at close orbital distances. The first exoplanet detected around a main-sequence star, 51 Peg b, is an example of a so-called Hot Jupiter, orbiting its host star in only 4.2~days \citep{Mayor1995}. Exoplanets in close orbits are subject to much higher levels of irradiation from the host star than any planets in our own solar system. The intense irradiation across the electromagnetic spectrum can cause inflated radii of Hot Jupiters (see \citet{Fortney2010, Baraffe2010} for reviews). In the UV and X-ray part of the spectrum, the stellar photons are absorbed at high altitudes in the exoplanetary atmosphere, where they can power a hydrodynamic evaporation process \citep{Watson1981, Murray-Clay2009}. Extended exoplanetary atmospheres as well as ongoing atmospheric escape have been detected through different observational setups that target certain parts of the spectrum where even optically thin atmospheric layers can cause enough absorption of starlight to produce observational effects during exo\-planetary transits. Examples are the Lyman-$\alpha$ line of hydrogen \citep{Vidal-Madjar2003, Lecavelier2010, Kulow2014, Ehrenreich2015}, the near-infrared metastable lines of helium \citep{Spake2018, Nortmann2018}, observations in the near-ultraviolet \citep{Salz2019}, and in soft X-rays \citep{Poppenhaeger2013}.

The main driver for exoplanetary atmospheric escape is thought to be the extreme-ultraviolet (EUV) and soft X-ray flux that the planet receives from the host star \citep{Yelle2004, Murray-Clay2009}. The EUV component of the stellar spectrum is currently not directly observable, because no space observatories with sensitivity at the corresponding wavelengths are in operation; the EUVE satellite was the last major EUV observatory and ceased operations in 2001. In contrast, the X-ray part of the stellar spectrum is observable with a variety of currently operating instruments. The stellar EUV flux, in turn, can be estimated from the stellar X-ray emission and the UV part of the stellar spectrum \citep{Sanz-Forcada2011, France2013}. Several uncertainties still exist when trying to estimate the mass-loss rates of exoplanets, for example the X-ray absorption height in exoplanetary atmospheres and the overall efficiency of exoplanetary mass loss \citep{Owen2014, Cohen2015, Dong2017}. However, one of the most important input quantities of exoplanet evaporation rates, namely the exoplanetary high-energy irradiation, can be determined through X-ray observations.

Launched in 2019, eROSITA is producing the first all-sky survey in X-rays since the ROSAT mission in the 1990s. In this work we present a catalogue of exoplanet X-ray irradiation levels derived from eROSITA's full-sky survey data and the eROSITA Final Equatorial Depth Survey (eFEDS) (see publication by Brunner et al.\ in this volume), 
augmented by archival observations from ROSAT, \textit{XMM-Newton}, and \textit{Chandra}. We calculate the stellar combined X-ray and EUV (in short, XUV) fluxes as well as estimates for the exoplanetary evaporation rates. We report on several exoplanets that are strongly irradiated in the high-energy regime, making them good candidates for observing ongoing evaporation signatures at other wavelengths.

The paper is structured as follows: Section~\ref{observations} describes the observations and data reduction; section~\ref{dataanalysis} describes the considerations used for catalogue matching and analysis performed to extract flux estimates for stellar coronae; section~\ref{results} gives the main results with respect to stellar X-ray fluxes and luminosities, exoplanetary irradiation levels, and mass loss rates; section~\ref{discussion} puts the results into the context of exoplanet evaporation; and section~\ref{conclusion} summarises our findings.

\section{Observations}\label{observations}

\subsection{eROSITA}

The eROSITA instrument consists of seven X-ray telescopes and CCD cameras onboard the Russian-German Spectrum-X-Gamma (SRG) spacecraft \citep{Sunyaev2021} and was launched into orbit in summer 2019. A detailed description of eROSITA is given in 
\citet{Merloni2012} and \citet{Predehl2021}. 
In short, eROSITA has a circular field of view with a diameter of 1.03$^\circ$, an average spatial resolution of 26$^{\prime\prime}$, and is sensitive to photons from an energy range of 0.2-10 keV. 
eROSITA observes the whole sky once within 6 months by scanning along great circles in the sky that are approximately perpendicular to the ecliptic, similar to the ROSAT All-Sky Survey \citep{Voges1999, Voges2000, Boller2016}. The survey portion of the eROSITA mission, called the eROSITA All-Sky Survey (eRASS), will last four years, in which the whole sky is scanned eight times.
Prior to starting the eRASS, eROSITA performed a Calibration and Performance 
Verification phase (CalPV), in which it observed an equatorial field of about 140\,deg$^2$ size for an average exposure time of ca.\ 2\,ks per pixel, in order to image a small patch of the sky to the same depth expected at the end of the 4-years all-sky survey. This eROSITA Final Equatorial Depth Survey (eFEDS) \citep{Brunner2021efeds} will be included in the Early Data Release of the eROSITA consortium in 2021.

We use data from the intermediate consortium-wide data release of the eRASS1 and eRASS2 surveys, meaning the first and second full-sky surveys performed by eROSITA. 
We have access to all eRASS X-ray sources located in the half of the sky which is proprietary to eROSITA\_DE, the German eROSITA collaboration
(i.e.\ with a galactic longitude larger than 180$^\circ$).
The raw data was processed with a calibration pipeline based on the eROSITA Science Analysis Software System (eSASS) (see Brunner et al.,\ submitted). The intermediate eRASS1 and eRASS2 catalogues give, among other parameters, the positions, detection likelihoods, and vignetting-corrected count rates of the detected X-ray sources in three energy bands, 0.2-0.6 keV, 0.6-2.3 keV, and 2.3-5.0 keV. Typical vignetting-corrected exposure times over an individual half-year survey are of the order of 150 seconds per source, but can differ strongly depending on the position of the source on the sky, with larger exposure times towards the ecliptic poles.

For stellar coronae, significant X-ray emission is typically found at energies below 5 keV, with the exception of extremely powerful (but transient) flares (see \citet{Guedel2004} for a review). We therefore concentrate our study on the three canonically extracted energy bands (0.2-0.6, 0.6-2.3 and 2.3-5.0 keV) of the intermediate eRASS catalogues.

\subsection{ROSAT}

ROSAT was a space telescope that observed the sky in soft X-rays in an energy range of 0.1-2.4 keV \citep{Truemper1982}, with an all-sky survey (RASS) as well as pointed observations. We use the Second ROSAT all-sky survey (2RXS) source catalogue from \citet{Boller2016}, which is available through the VizieR service. To obtain stellar coronal fluxes, we use the counts-to-flux conversion formula from \citet{Schmitt1995}, which uses detected count rates and hardness ratios from RASS for a flux calculation. We later scale these fluxes to a canonical energy band of 0.2-2 keV, with details given in subsection~\ref{fluxconversion}.

\subsection{XMM-Newton}\label{xmm}

\textit{XMM-Newton} is an X-ray mission with several instruments on board \citep{Jansen2001}. Relevant for our analysis here is only the data collected by the EPIC instrument, consisting of three CCD cameras \citep{Turner2001, Strueder2001}. The energy range and spatial resolution of EPIC is similar to eROSITA. 
The \textit{XMM-Newton} mission provides a number of different source catalogues, including merged source detections from pointed observations, the slew survey, and multiply-observed sources
 (see for example \citealt{Saxton2008, Watson2009, Traulsen2020}). We make use of the 4XMM-DR10 catalogue\footnote{\url{http://xmmssc.irap.omp.eu/Catalogue/4XMM-DR10/4XMM_DR10.html}} in its ''slim'' version, where the longest existing exposure has been selected for any given source.

\subsection{Chandra}

\textit{Chandra} is an X-ray telescope with two X-ray imaging instruments, ACIS and HRC \citep{Weisskopf2002, Garmire2003, Murray1997}. HRC is sensitive to photon energies from 0.08-10.0 keV, but provides no intrinsic energy resolution. ACIS has an intrinsic energy resolution of 50 eV (FWHM) at soft energies, and has an energy sensitivity of 0.2-10.0 or 0.6-10.0 keV, depending on which chip of the ACIS instrument a source falls onto. We used the \textit{Chandra} Source Catalog (CSC), Release 2.0 \citep{Evans2010, Evans2018} for our analysis, which is available through the VizieR interface.

\section{Data Analysis}\label{dataanalysis}

\subsection{Catalogue cross-matching}

\begin{figure}
\includegraphics[width=0.5\textwidth]{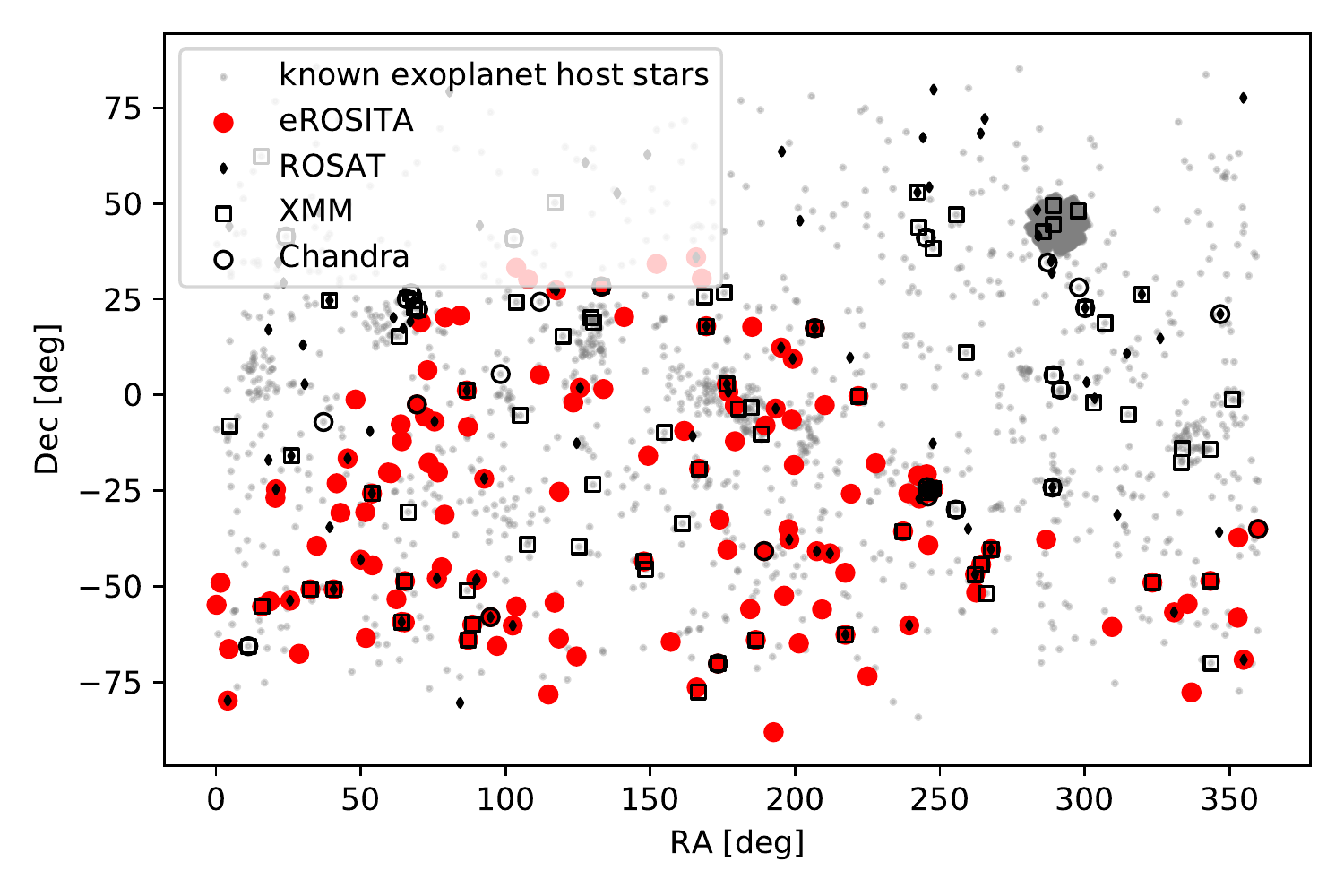}
\caption{X-ray detections of known exoplanet host stars in the sky. Known planet host stars are depicted as small grey dots; the Kepler field at RA $=$ 300 deg as well as the increased density of known planets along the ecliptic due to the coverage by the K2 mission are visible. Detections in the German eROSITA sky with the eRASS1 or eRASS2 survey are shown as red filled circles, previous detections with ROSAT, \textit{XMM-Newton} and \textit{Chandra} are shown as black filled diamonds, open circles and open squares, respectively.}
\label{fig:sky}
\end{figure}

We used the NASA Exoplanet Archive's catalogue of detected exoplanets as our starting point. We downloaded the full table of confirmed exoplanets and their properties on March 26, 2021 \footnote{\url{https://exoplanetarchive.ipac.caltech.edu}}, using their default data sets for each exoplanet. We excluded the small number of exoplanets detected by the microlensing method, because their stellar distances have large uncertainties of the order of 50\%, which would propagate into our final exoplanetary mass loss rates as very large uncertainties. We also discarded one entry in the exoplanet table, namely the postulated exoplanet around the cataclysmic variable HU Aqr, because the planet has been shown to be spurious \citep{Schwope2014, Bours2014, Gozdziewski2015}. We plot these remaining exoplanets from the catalogue as the grey points in Fig.~\ref{fig:sky}.

The host star coordinates in the Exoplanet Archive table are based on optical observations and are given by NASA for epoch J2015.5 for all sources with a \textit{Gaia} DR2 source ID \citep{gaiadr2, Lindegren2018}, and for epoch J2000 for the remaining few exoplanet host stars without a \textit{Gaia} DR2 entry. We propagated all host star coordinates to epochs suitable for catalogue matching with the respective X-ray catalogues, using the \textit{Gaia} DR2 proper motions where available, and \textsc{Hipparcos} proper motions otherwise. Typical proper motions of known exoplanet host stars within a distance of 100\,pc from the Sun are of the order of 200\,$\mu$arcsec/yr and significantly smaller at larger distances, but a small number of stars in the sample display proper motions upwards of 1\,arcsec/yr.

We then performed a positional source matching of the exo\-planet catalogue with the individual X-ray catalogues. The closest X-ray source in a chosen matching radius to an exoplanet host star was selected as the fiducial match. Maximum matching radii were based on considerations of both the typical positional uncertainties of the respective telescopes and the expected uncertainties in propagated stellar positions at the observing epoch. The typical positional uncertainties of the X-ray catalogues are of the order of 12.5$^{\prime\prime}$ for ROSAT \citep{Voges1999}, 1.6$^{\prime\prime}$ for \textit{XMM-Newton}\footnote{\url{https://www.cosmos.esa.int/web/xmm-newton/news-20201210}}, and 0.8$^{\prime\prime}$ for \textit{Chandra}\footnote{\url{https://cxc.harvard.edu/cal/ASPECT/celmon/}}.
For eROSITA, the current positional uncertainty in the existing data reduction version is of the order of 5$^{\prime\prime}$, however, this is expected to improve with further detailed analysis and re-reduction of the data. As ROSAT's RASS survey and the eROSITA eRASS surveys span only narrow epoch ranges, we opted for maximum matching radii of twice the typical positional uncertainties of those catalogues, after propagating the stellar positions to an epoch of J1990 for RASS and J2020.25/J2020.75 for eRASS1/eRASS2, respectively. For \textit{XMM-Newton} and \textit{Chandra}, however, their obs\-erving epochs span a range of roughly 20 years each, in which significant motions of some of our sample stars can accrue. We therefore initially matched the stars to the \textit{XMM-Newton} and \textit{Chandra} catalogues with large matching radii of 30$^{\prime\prime}$, determined the observational X-ray epoch from the preliminary matches, and then performed a second source matching with suitably propagated stellar positions and narrower maximum matching radii of 5$^{\prime\prime}$ for \textit{XMM-Newton} and 2$^{\prime\prime}$ for \textit{Chandra}\footnote{In cases where the observed co-added epochs in the catalogues spanned more than three years and proper motions were large or not available, the maximum allowed matching radius was doubled.}.

The exoplanet host star catalogue is a sparse catalogue containing about 3200 stars over the whole sky, compared to about 700,000 X-ray sources in the eRASS1 catalogue covering the German half of the sky. The other used X-ray catalogues are denser than the exoplanet host star catalogue as well. It is therefore not surprising that we do not find any true double matches in our proximity-based matching. The only double match, where more than one entry in the host star catalogue was matched to the same X-ray source in eRASS and ROSAT, was for the system HD~41004AB. In this system two stars with an on-sky separation of about 0.5$^{\prime\prime}$ are both orbited by known exoplanets, with the lower-mass star being positioned about 0.5$^{\prime\prime}$ to the south of the primary \citep{Raghavan2010}. The system was also observed with \textit{Chandra}, where visual inspection shows an X-ray bright source at the position of the B component, and no additional X-ray source visible at the position of the A component. We therefore attributed all X-ray flux stemming from the HD~41004AB system to component B.

While there are no further double matches among the catalogues matched here, it is known that several exoplanet host stars are common proper motion binaries with other cool stars \citep{Raghavan2010, Mugrauer2019} that are X-ray sources as well \citep{Poppenhaeger2014}. Often these companion stars are not known to host an exoplanet themselves and are therefore not listed in the exoplanet host star catalogue. Some of the companion stars are close enough to the planet host stars to not be spatially resolved by some of the used X-ray telescopes. In such cases, we split the X-ray flux stemming from the system equally between the unresolved stellar components. A more detailed analysis of such systems will be presented in Ilic et al.\ (in prep.); the list of stars where such a split was performed in this work is given in the Appendix.

\begin{figure}
\includegraphics[width=0.5\textwidth]{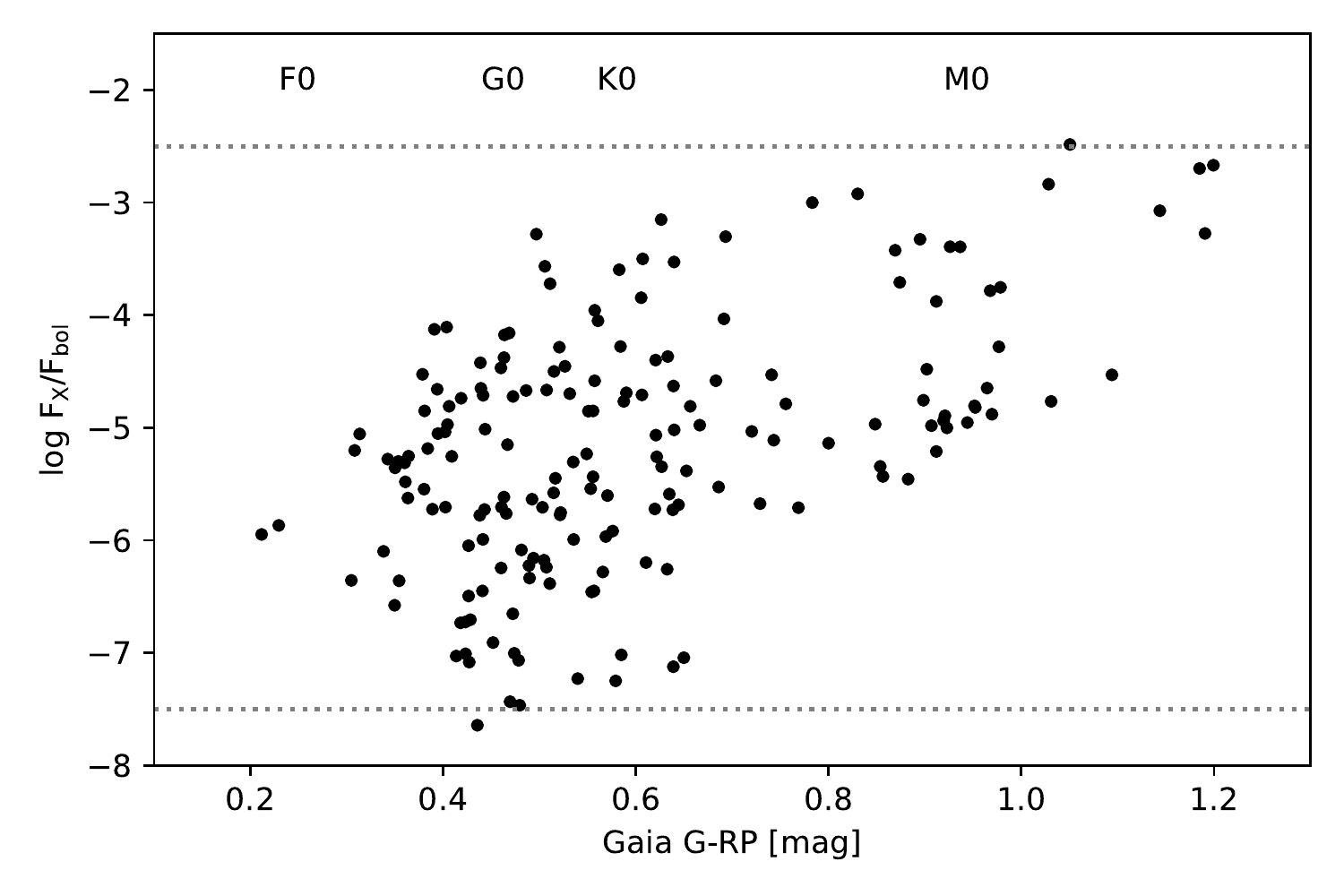}
\caption{X-ray to bolometric flux ratios of the exoplanet host stars in our sample versus their \textit{Gaia} colour $G-R_{p}$; corresponding spectral types are given at the top of the figure. The horizontal dotted lines indicate the approximate upper and lower boundaries of typically observed flux ratios for main-sequence stars, which our sample agrees with well.}
\label{fig:Rx}
\end{figure}

To test whether our fiducial X-ray matches can be accepted as bona fide counterparts to the exoplanet host stars, we analysed the ratio of X-ray to bolometric flux for the fiducial matches. Stellar coronae are known to exhibit a ratio of $\log F_X/F_{bol}$ between -2.5 and -7.5 for most stars. Astrophysical exceptions are flaring low-mass stars which can temporarily display values of up to -2 and stars with extremely low or no magnetic activity, such as Maunder minimum stars in the former case or stars with masses high enough to prohibit an outer convective envelope in the latter. We extracted bolometric fluxes for exoplanet host stars with \textit{Gaia} DR2 source IDs directly from the \textit{Gaia} DR2 archive where bolometric luminosities were derived with the FLAMES algorithm 
\citep{Andrae2018}, which yielded $L_\mathrm{bol}$ values for 184 out of 241 X-ray detected host stars.
After unifying X-ray fluxes from different catalogues for a stellar coronal spectral model and a common energy band as described in section~\ref{fluxconversion}, we found that the distribution of the X-ray to bolometric flux ratio of our matched sources is well within expectations for stellar coronal sources (Fig.~\ref{fig:Rx}).

\begin{figure}
\includegraphics[width=0.5\textwidth]{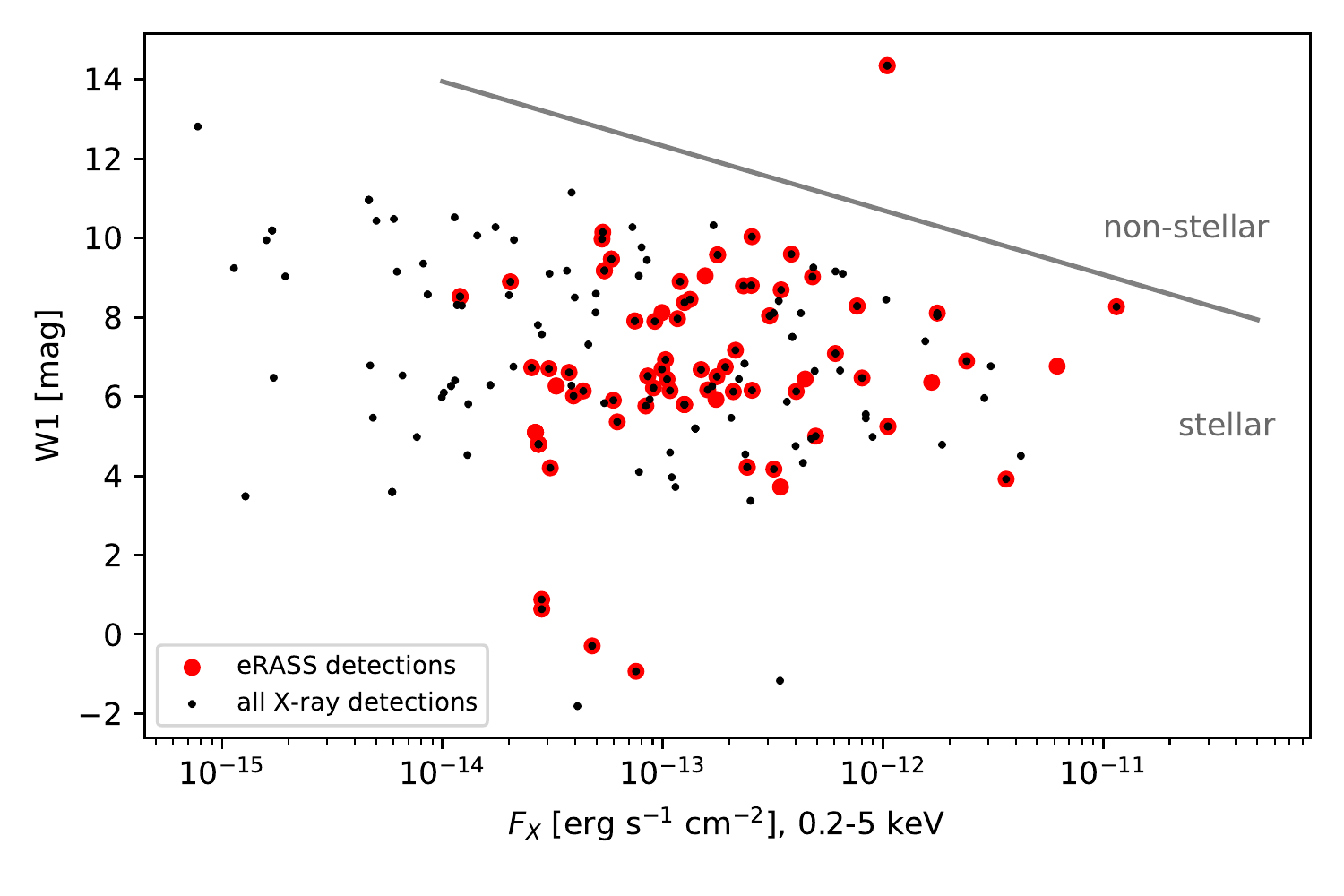}
\caption{Fluxes of the X-ray detected stars in our sample in the soft X-ray band and the WISE W1 infrared band, large red dots for eRASS detections and small black dots for all X-ray detections. The statistical dividing line between objects of stellar and non-stellar nature \citep{Salvato2018} is shown as the grey solid line. The only source that falls into the non-stellar part of the parameter space is an exoplanet-hosting cataclysmic variable.}
\label{fig:starnonstar}
\end{figure}

Furthermore, we compared the soft X-ray fluxes to the infrared fluxes of the matched targets. \citet{Salvato2018} found that stars typically display higher infrared brightness in the WISE W1 band for a given soft X-ray flux than non-stellar X-ray sources such as Active Galactic Nuclei (AGN), with stellar and non-stellar objects being well separated in a plane spanned by the X-ray flux and the W1 magnitude.
We display our matched X-ray and optical sources, the majority of which have known W1 magnitudes listed in the exoplanet catalogue, in Fig.~\ref{fig:starnonstar}. Almost all of our matched sources fall into the stellar area of the diagram; the single source that falls into the non-stellar part of the diagram is an exoplanet-hosting object that is not a main-sequence star, namely the cataclysmic variable UZ~For. We therefore consider our catalogue matches to be unlikely to be contaminated by extragalactic sources.

\subsection{Flux conversions}\label{fluxconversion}

The used X-ray catalogues provide fluxes in slightly different energy bands. We opted for a commonly used soft X-ray band of 0.2-2~keV for the analysis of X-ray irradiation levels of exoplanets. We describe in the following how any necessary conversion factors were derived.

The intermediate eRASS catalogues use, similarly to the \textit{XMM-Newton} catalogue, an assumed absorbed powerlaw spectral model to calculate X-ray fluxes from count rates. The assumed underlying model has an absorption column of $N_H = 1\times 10^{20}$ and a powerlaw index of 1.7 for this intermediate version of the eRASS catalogues. 

This is not a suitable spectral model for stellar coronae, which are described by an optically thin thermal plasma, with a contribution from absorption by the interstellar medium that tends to be small since detected exoplanets are typically located close to the Sun (with more than 80\% of the currently detected exoplanet host stars being located within a distance within 100 pc from the Sun).

\begin{figure}
\includegraphics[width=0.5\textwidth]{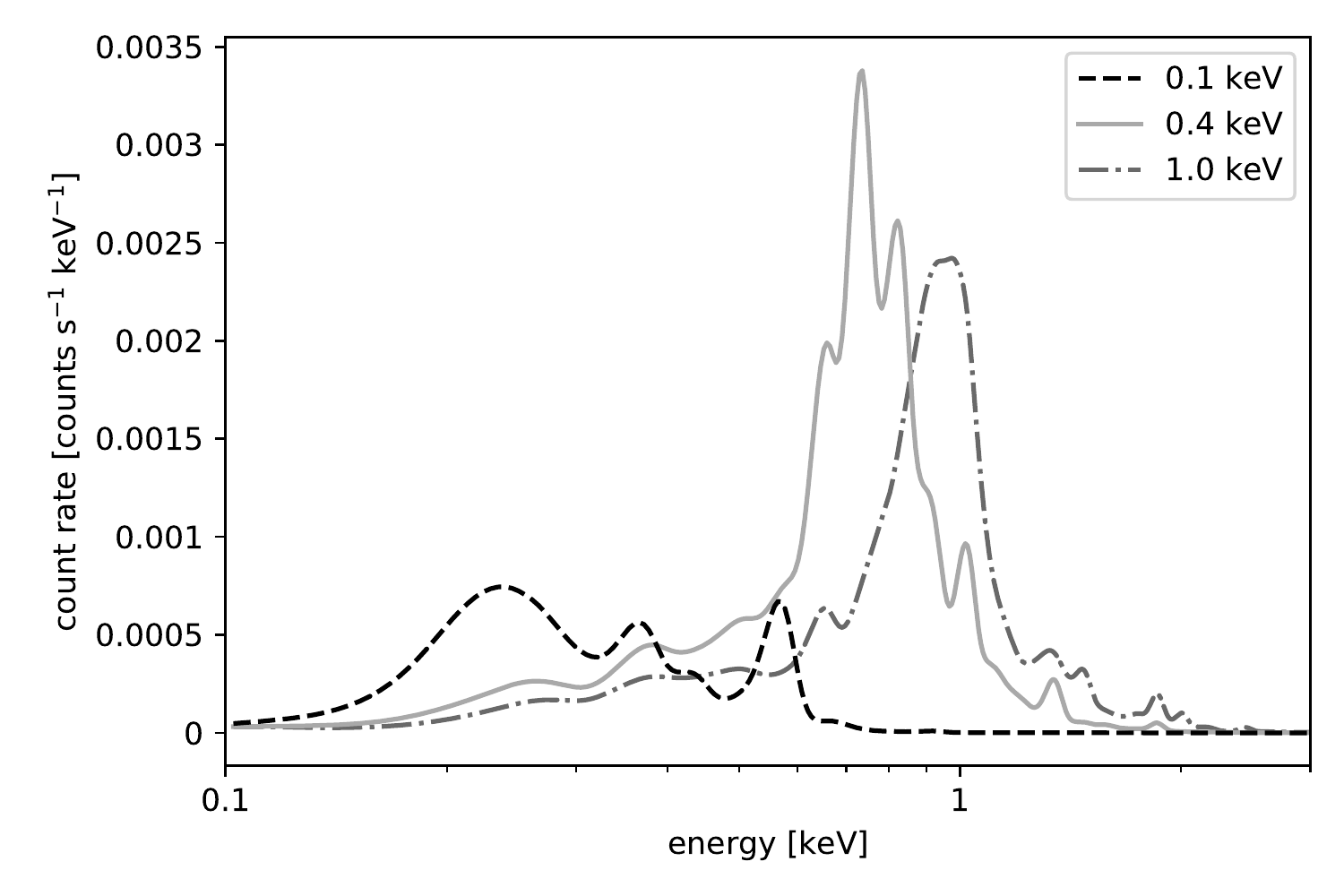}
\caption{Simulated stellar coronal spectra, using eROSITA's instrumental response. A single-temperature coronal plasma model was used with temperatures of 0.1, 0.4 and 1 keV (1.1, 4.6 and 11.4 million K). Even for very hot stellar coronae almost all photons are emitted at energies below 2 keV.}
\label{fig:spectra}
\end{figure}

\begin{figure}
\includegraphics[width=0.5\textwidth]{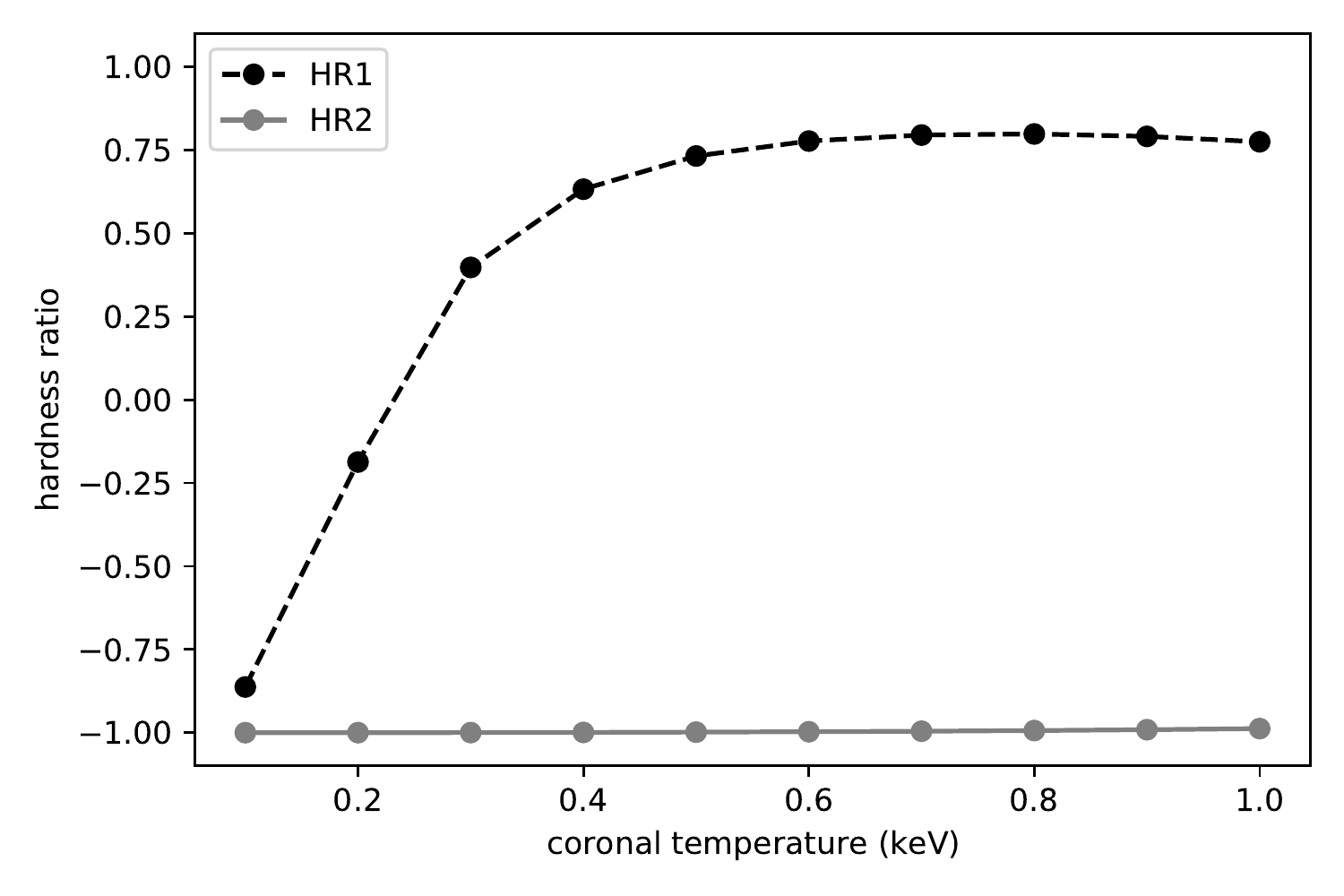}
\caption{Hardness ratios HR1 and HR2 from simulated stellar coronal eROSITA spectra, as a function of coronal temperature. HR2 is always close to -1, rising only very slightly for very high temperatures, while HR1 rapidly rises from low to moderate coronal temperatures and then saturates at a value of about 0.75 for temperatures above 0.5 keV.}
\label{fig:HR_vs_temp}
\end{figure}

\begin{figure}
\includegraphics[width=0.5\textwidth]{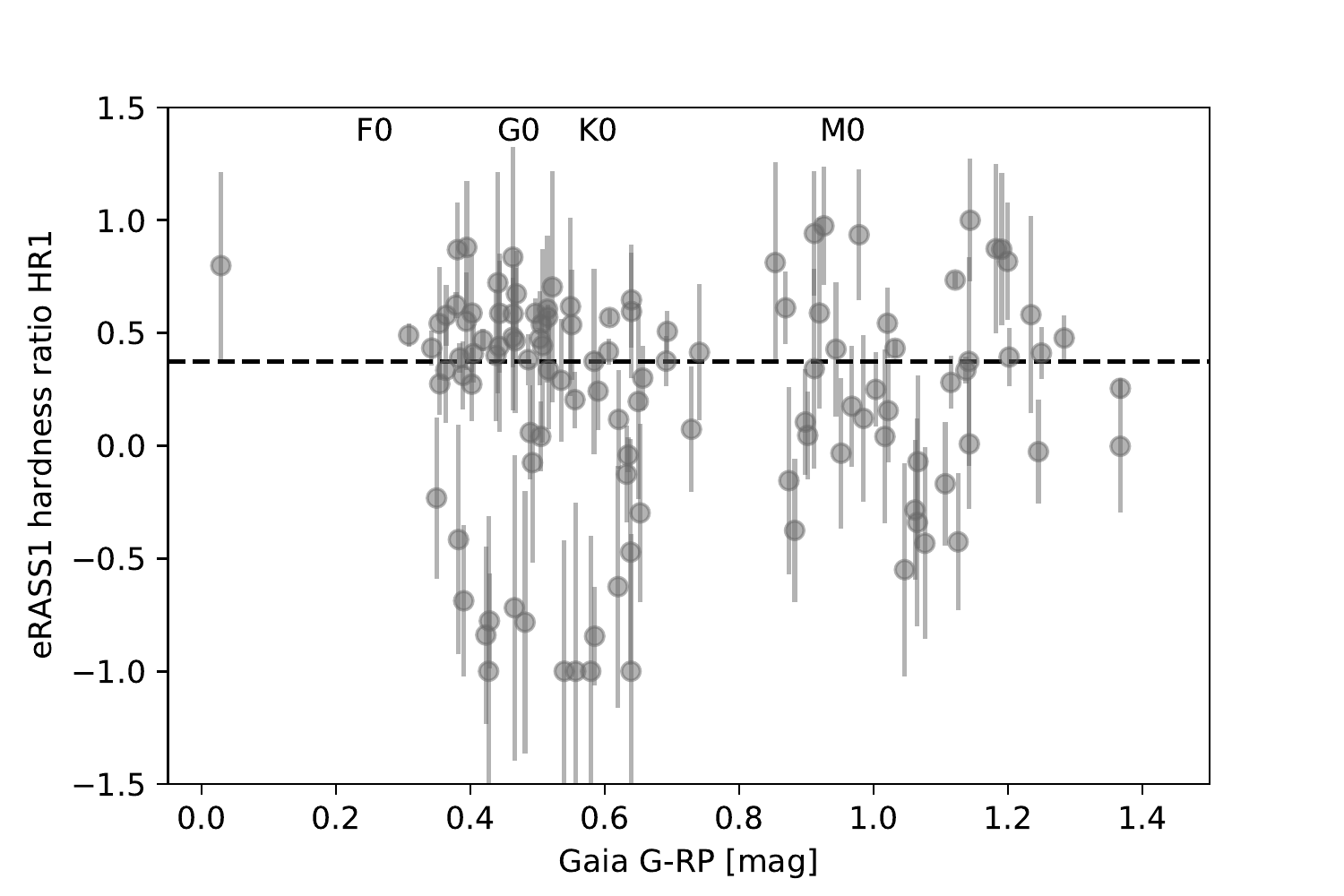}
\caption{Observed hardness ratio HR1 with 1$\sigma$ uncertainties versus the \textit{Gaia} $G-RP$ colour for the stars in our sample. The median hardness ratio, corresponding to a coronal temperature of about 0.3 keV, is depicted by the dashed line. The quite blue star to the left is the Herbig Be star HD~100546, a known X-ray emitter \citep{Skinner2020}.
}
\label{fig:hr}
\end{figure}

In order to test if an assumption of a typical coronal temperature of 0.3 keV is appropriate for the eROSITA-detected planet host stars, 
we first performed an analysis of X-ray hardness ratios in relation to coronal temperature. We simulated eROSITA spectra with Xspec version 12.11.1 \citep{Arnaud1996} using the eROSITA instrumental response files. Because the eRASS surveys are currently still shallow, we omitted an absorbing column and simulated spectra with a single temperature component for a range of coronal temperature parameters with $kT$ between 0.1 and 1.0 keV steps, corresponding to temperatures of ca.\ 1.1 to 11 million K; we show some of those spectra in Fig.~\ref{fig:spectra}. 
For each of those simulated spectra we calculated their model-based fluxes in the 0.2-0.6 (S), 0.6-2.3 (M) and 2.3-5.0 (H) keV energy bands, as well as their simulated count rates in those bands and the corresponding hardness ratios HR1 = (M-S)/(M+S) and HR2 = (H-M)/(H+M). We find that for the typical range of coronal temperature simulated by us, the hardness ratio HR2 is always very close to -1, which is due to the fact that the effective area of eROSITA drops significantly beyond 2.3 keV. The simulated softer hardness ratio HR1 ranges from -0.9 to 0.8. We display the relationship between the modelled coronal temperature and the simulated eROSITA hardness ratios in Fig.~\ref{fig:HR_vs_temp}. Note that this is valid for stars whose coronal spectra are dominated by a single temperature component; stars multiple and strongly different coronal temperature components can behave differently with respect to their observed hardness ratios.
In the observed data for our sample stars, the hardness ratio HR1 spans the full range between -1 and 1, with typical uncertainties of about $\pm$0.2, i.e.\ consistent with the simulated range of values. The majority of our observed stars concentrates between values from -0.1 to 0.9 (Fig.~\ref{fig:hr}), with a median of 0.34. This corresponds to a coronal temperature of about 0.3~keV. The observed values of HR2 are in agreement with a value of -1 within observational uncertainties.

Given the observed hardness ratios, the eROSITA-detected sample is in good agreement with a typical coronal temperature of 0.3~keV, which we use to correct the final fluxes from a power-law to a stellar coronal model. We find the conversion factor between the fluxes to be $F_\mathrm{X,\,coronal} = 0.85 F_\mathrm{X,\,powerlaw}$.  The eRASS stellar fluxes were calculated by applying the relative conversion factor to the powerlaw-derived fluxes from the intermediate eRASS catalogues.

For ROSAT, we again use a typical coronal temperature of $kT=0.3$\,keV to transform the fluxes from the 0.1-2.4 keV band to the canonical 0.2-2 keV band, using the WebPIMMS tool. The found conversion factor is $F_\mathrm{X, 0.2-2\,keV} = 0.87 \times F_\mathrm{X, 0.1-2.4\,keV}$.

For \textit{XMM-Newton}, the 0.2-2 keV band already is one of the canonical bands given in the source catalogues, as the sum of bands 1 (0.2-0.5 keV), 2 (0.5-1.0 keV) and 3 (1.0-2.0 keV). However, since the \textit{XMM-Newton} catalogue fluxes assume an underlying powerlaw spectrum with with $N_H = 3\times 10^{20}$ and powerlaw index of 1.7 \citet{Rosen2016}, we need to correct these fluxes to an underlying stellar coronal model. We again choose as a representative stellar model a coronal plasma with a temperature of 0.3 keV. \textit{XMM-Newton} typically has deep pointings, detecting sources at larger distances than eROSITA, so for this model a non-zero absorption column of $N_H = 3\times 10^{19}$ was used.
The relative corrections compared to the power-law fluxes depend on the specific instruments and filters used in a given observations. Using the WebPIMMS tool\footnote{\url{https://heasarc.gsfc.nasa.gov/cgi-bin/Tools/w3pimms/w3pimms.pl}}, we find a typical correction factor of $F_\mathrm{X,\,coronal} = 0.87 F_\mathrm{X,\,powerlaw}$ for the 0.2-2 keV band for the combined signal from the EPIC cameras.

For \textit{Chandra}, its second source catalogue also lists fluxes that are not explicitly model-dependent, which are derived from the energies of the detected photons and the effective area of the instrument at those energies. We construct the soft flux in the 0.2-2.0 keV band by combining the $u$ (0.2-0.5 keV), $s$ (0.5-1.2 keV), and $m$ (1.2-2.0 keV) bands. However, it needs to be noted that depending on the instrument used in a given observation the flux in the softest band may have gone undetected, because the ACIS-I configuration has a very small effective area at the softest energies. Therefore, some of the \textit{Chandra}-derived fluxes may underestimate the true soft-band X-ray flux of a star.

\subsection{Optical loading in eROSITA data}

Objects with high optical brightness can cause spurious signals in X-ray observations. While X-ray CCDs are mainly sensitive to genuine X-ray photons, a large number of optical and infrared photons impinging on a CCD pixel within a readout time frame can release electrons in the CCD, which can be falsely attributed to an X-ray photon event; this is called ''optical loading''.

\begin{figure}
\includegraphics[width=0.5\textwidth]{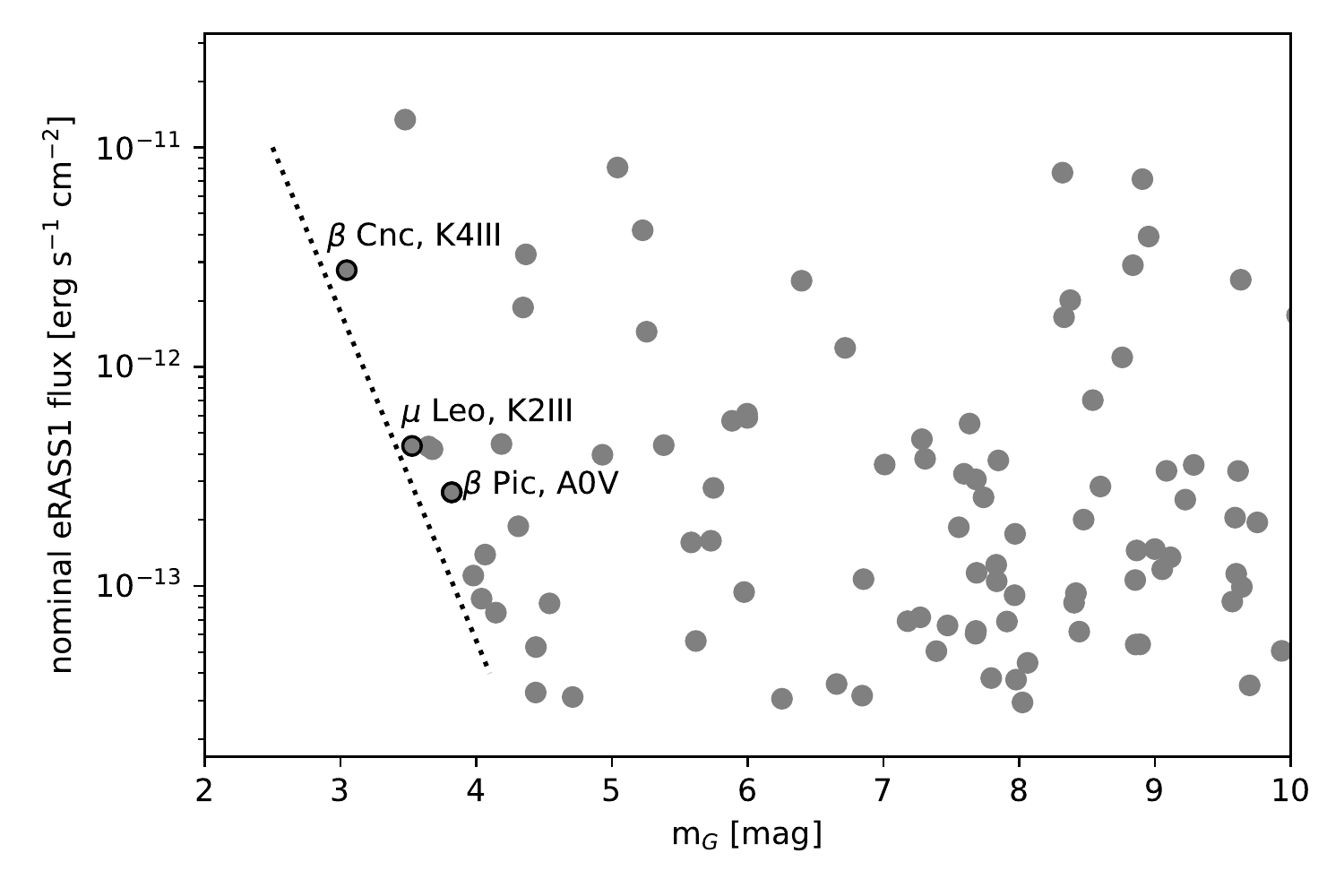}
\caption{Nominal X-ray fluxes from eRASS1 versus optical brightness for stars in our sample. For stars with an apparent \textit{Gaia} magnitude brighter than 4 mag there is a clear trend towards high apparent eRASS fluxes, which can be attributed to optical loading. Some individual bright stars named in the plot can be expected to be only weak X-ray emitters, because they are either lacking an outer convective envelope ($\beta$~Pic) or are ''coronal graveyard''-type giants \citep{Ayres2003}. Furthermore, these specific stars are known to be X-ray dim from previous observations by other X-ray telescopes.}
\label{fig:opticalloading}
\end{figure}

We show the nominal X-ray fluxes from the eRASS1 catalogue versus the optical \textit{Gaia} magnitude of host stars in our sample in Fig.~\ref{fig:opticalloading}. 
We find that stars with an optical brightness of $m_G = 4$\,mag or brighter in the \textit{Gaia} band display an apparent floor to their detected eRASS fluxes which rises with optical brightness. The A0V star $\beta$~Pic is one of those stars, and it is known to be X-ray dimmer by two orders of magnitude from previous pointed X-ray observations \citep{Hempel2005, Guenther2012}. We therefore attribute the apparent X-ray flux of optically bright sources with $m_G \lesssim 4$\,mag to optical loading in eROSITA observations and discard their contaminated eROSITA X-ray fluxes from the further analysis. We note that at least one optically bright star, $\epsilon$~Eridani, is a genuinely X-ray bright star which is known from observations with other X-ray telescopes \citep{Poppenhaeger2010, Coffaro2020}. However, a detailed spectral analysis of the eROSITA data to tease apart its coronal X-ray emission and the optical loading is beyond the scope of this work. We therefore use $\epsilon$~Eri's measured X-ray flux from \textit{XMM-Newton} in the further analysis.

\section{Results}\label{results}

\subsection{New X-ray detections of exoplanet host stars}

\begin{figure}
\includegraphics[width=0.5\textwidth]{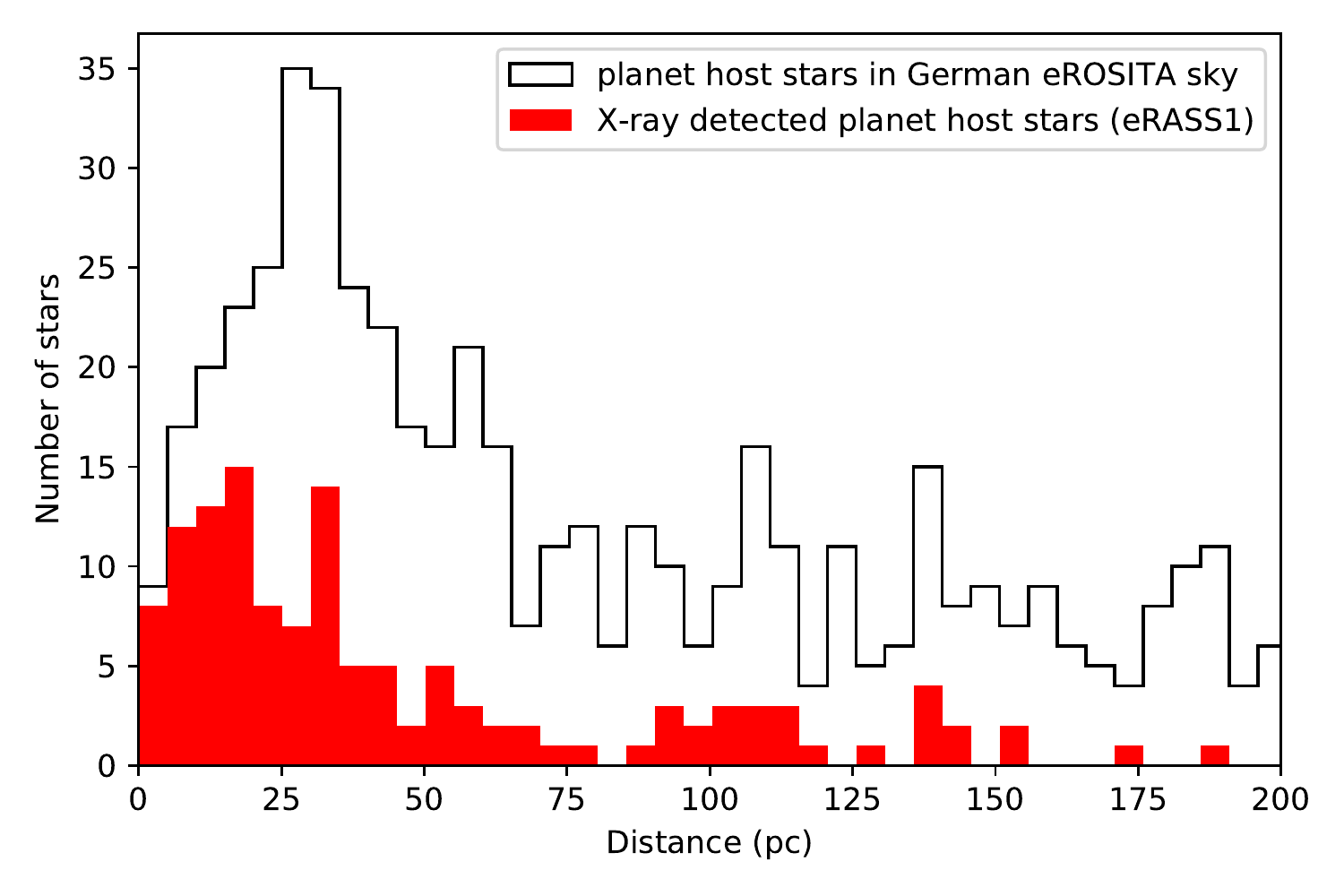}
\caption{Histogram of exoplanet host stars in distance bins of 5\,pc (white with black outline) and the exoplanet host stars detected in the eRASS1 survey (red) in the German eROSITA sky, out to a distance of 200\,pc. The detection fraction is high with ca.\ 70\% out to 20\,pc and then drops off rapidly. A small number of X-ray detections exists for planet host stars at larger distances.}
\label{fig:detectionfraction}
\end{figure}

We show the positions of X-ray detected exoplanet host stars in the sky in Fig.~\ref{fig:sky}. The total number of X-ray detected planet-hosting stars increases from 164 in the pre-eROSITA epoch to 241, i.e.\ 77 are added through eRASS1 and eRASS2. This increase can be expected to roughly double with the data from the Russian half of the eROSITA data. The X-ray detection fraction of exoplanet host 
stars in the German eROSITA sky is 
89\% within a distance of 5\,pc and 
70\% within 20\,pc. At larger distances the detection fraction drops rapidly (Fig.~\ref{fig:detectionfraction}).

\subsection{Flux comparisons between different X-ray missions}

\begin{figure}
  \includegraphics[width=0.5\textwidth]{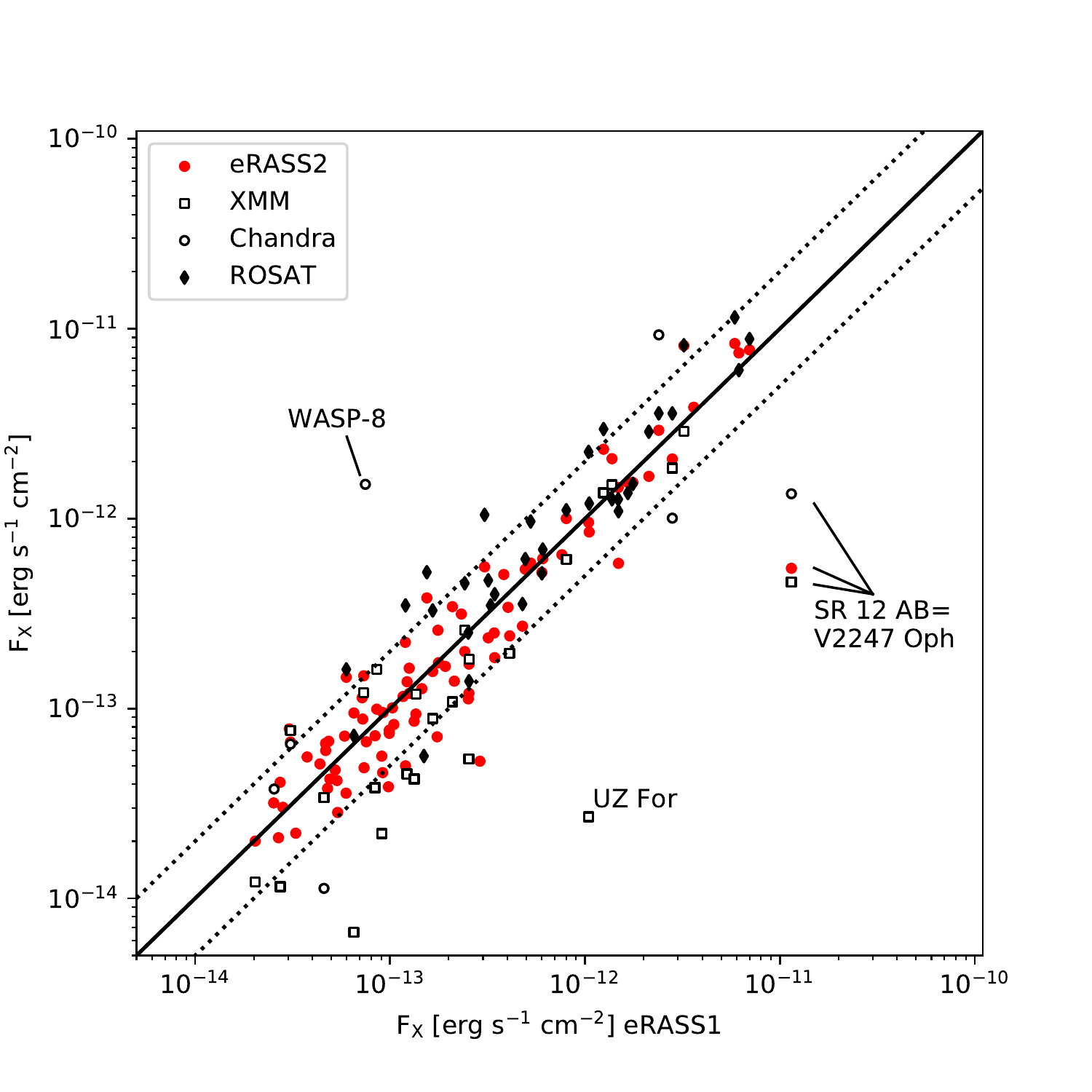}
\caption{Comparison of soft X-ray fluxes in the 0.2-2 keV band for exo\-planet host stars detected by more than one X-ray mission to fluxes detected by eRASS1. Stars detected by both eRASS1 and eRASS2 are shown in red. Some individual outliers are marked by name; they are objects expected to display strong variability in X-rays.}
\label{fig:fluxflux}
\end{figure}
After performing the flux conversions for all used data sets to cover the 0.2-2 keV range and correcting for an underlying stellar coronal model as laid out in section~\ref{fluxconversion}, we compare the fluxes observed for stars that were detected in more than one mission. We show the X-ray fluxes for stars that were detected in eRASS1 versus their fluxes observed in eRASS2, ROSAT, 
\textit{XMM-Newton} and \textit{Chandra} in Fig.~\ref{fig:fluxflux}. About 75\% of the observations display X-ray fluxes that agree within a corridor of 0.3 dex, which covers typical low-level intrinsic variability of stellar coronae. The nominal flux discrepancies grow larger towards the faint end, which is to be expected since also the flux uncertainties of the individual measurements increase. Additionally, when comparing data sets from different surveys and/or pointed observations, the shallowest survey will detect stars at the bottom of the survey's sensitivity only when they happen to be temporarily X-ray bright, for example because of a flare. We see this effect in two directions here: the ROSAT survey is shallower than eRASS, which is why we see the ROSAT-eRASS data-points skewing towards higher ROSAT fluxes in the X-ray faint regime. On the other hand, eRASS tends to be much shallower than pointed \textit{XMM-Newton} observations, which is why we see the XMM-eRASS data-points skewing towards higher eRASS fluxes at the X-ray dim end. For \textit{Chandra}, there are not enough common detections to cause any visible skew.

Some individual notable outliers in the plot stem from intrinsically strongly variable stars, such as the binary T~Tauri star V2247~Oph, the cataclysmic variable UZ~For, and the M dwarf GJ~176 which is known to flare frequently \citep{Loyd2018}.

\begin{figure}
\includegraphics[width=0.5\textwidth]{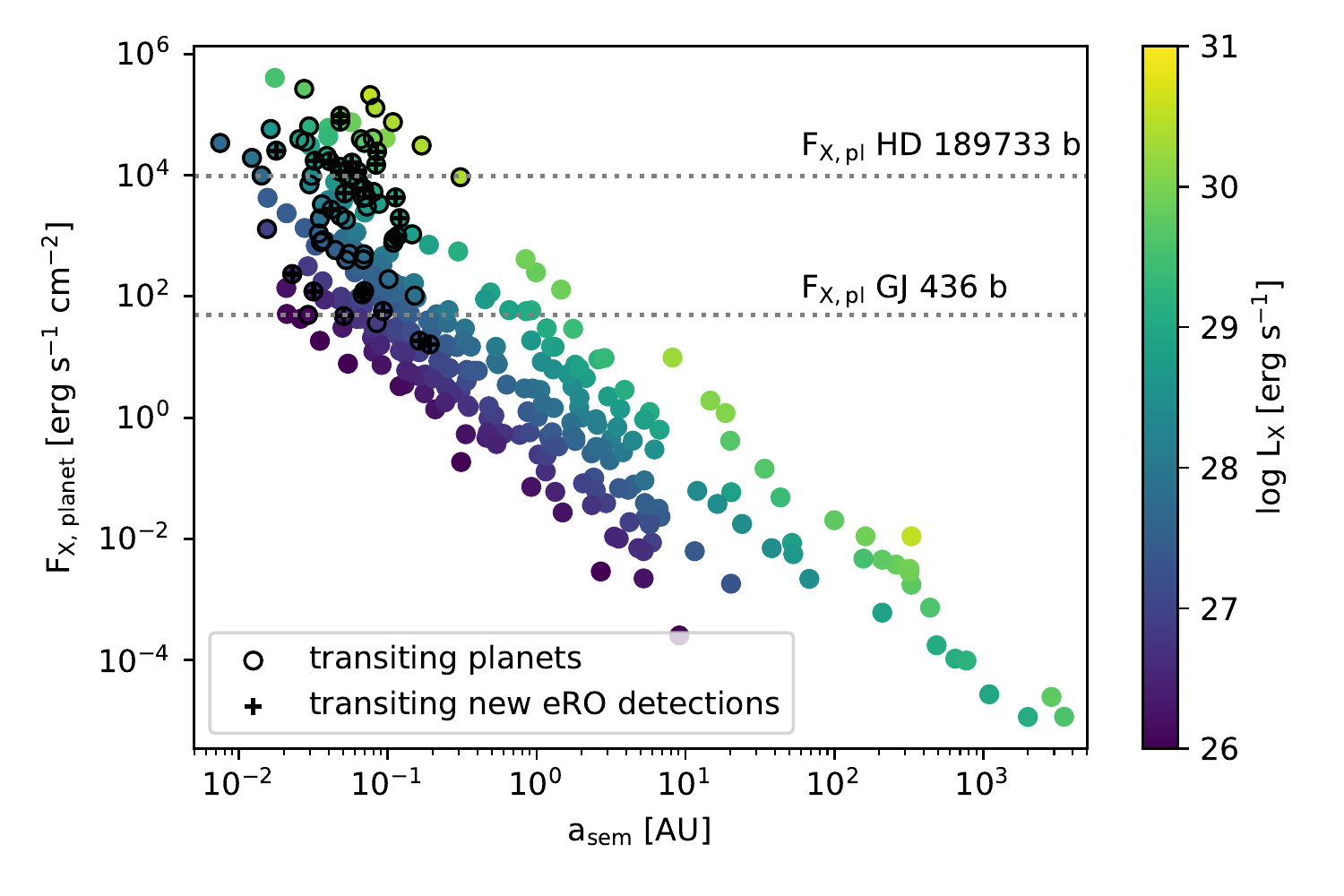}
\caption{X-ray irradiation fluxes of exoplanets versus their orbital semi-major axis. The vertical spread represents the intrinsic luminosity distribution of the host stars. Transiting exoplanets are marked with black open circles; new eROSITA X-ray detections among those are additionally marked with a black cross. For guidance, the X-ray irradiation fluxes of known evaporating exoplanets, the Hot Jupiter HD 189733 b and the warm Neptune GJ 436 b, are shown as horizontal dotted lines.
}
\label{fig:fxplanet}
\end{figure}

\subsection{X-ray irradiation and mass-loss of exoplanets}

Exoplanets experiencing an intense high-energy irradiation are expected to lose parts of their atmosphere through a so-called energy-limited escape process, which is much more efficient than Jeans escape \citep{Watson1981}. The incoming X-ray and extreme-UV flux (in short, XUV flux) is assumed to be the driver for this process. The process describes that a certain part of the impinging high-energy flux heats the upper layers of the exoplanetary atmosphere, which expands upwards and can push the layers above out of the gravitational well of the exoplanet. There are known limitations to the energy-limited escape model, for example for very high X-ray irradiation levels is it expected that hydrogen line cooling will start playing a more significant role, so that less energy is converted into atmospheric expansion \citep{Murray-Clay2009}. Also magnetic effects such as stellar winds interacting with the planetary atmosphere or a planetary magnetosphere shielding the planet from winds may play a role \citep{Owen2014, Cohen2015, Dong2017}. In the context of this work, we use a simple energy-limited hydrodynamic escape model based on \citet{Lopez2012, Owen2012}, with an atmospheric mass loss rate given by:

\begin{figure}
\includegraphics[width=0.5\textwidth]{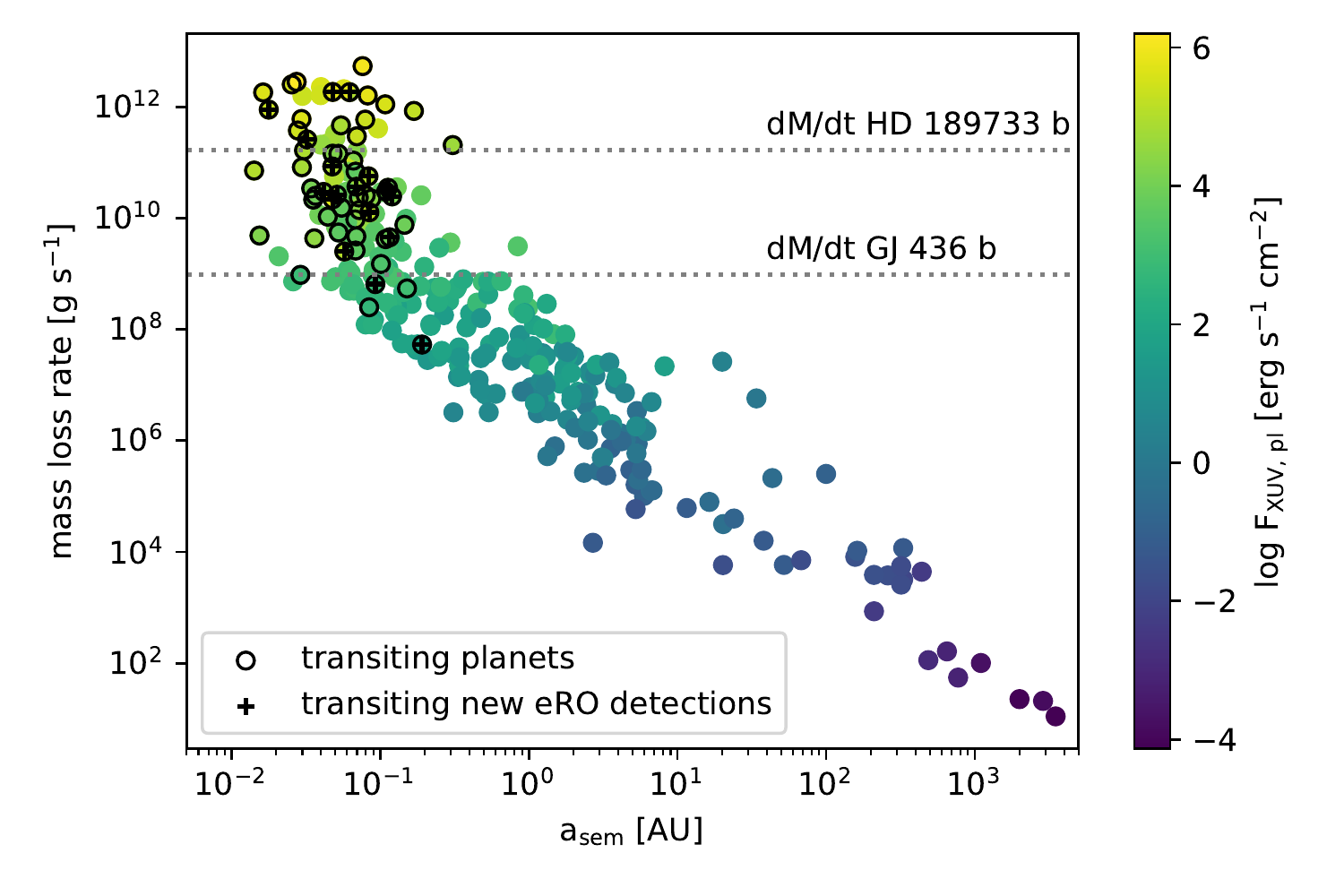}
\caption{Estimated mass loss rates of exoplanets under the energy-limited escape model (see text for details). The vertical spread represents the intrinsic luminosity distribution of the host stars. Transiting exoplanets, which are in principle accessible to follow-up observations to detect ongoing mass loss, are marked with black open circles; new eROSITA X-ray detections among those are additionally marked with a black cross. For guidance, the estimated mass loss rates of known evaporating exoplanets, the Hot Jupiter HD 189733 b and the warm Neptune GJ 436 b, are shown as horizontal dotted lines.}
\label{fig:mdotplanet}
\end{figure}

\begin{equation}
\dot{M} = {\epsilon}{{\pi R^{2}_{XUV}F_{XUV}}\over{ KGM_{pl}/R_{pl}}}
\end{equation}

where $\epsilon$ is the efficiency of the atmospheric escape, which we assume to be 0.15, $G$ is the gravitational constant, $K$ is a factor representing the impact of Roche lobe overflow, which we assume to be negligible for most systems and set to 1, $M_{pl}$ is the mass of the planet, $R_{pl}$ is the optical radius of the planet, $F_{XUV}$ is the X-ray and extreme-UV flux incident on the planet, and $R_{XUV}$ is the planetary radius at XUV wavelengths, which we assume to be 1.1 times the optical radius. For a detailed discussion of the assumptions made in this model we refer to \citet{Poppenhaeger2021}.

To estimate stellar XUV fluxes from X-ray fluxes alone, a variety of approaches exist \citep{Sanz-Forcada2011, Linsky2014, Chadney2015, Johnstone2020arXiv, King2021}. We choose the conversion relationship between X-ray and EUV fluxes by \citet{Sanz-Forcada2011}, which uses spectral energy distributions determined for a sample of cool stars. We first convert our X-ray fluxes for the 0.2-2~keV band into the required input band of 0.1-2.4\,keV for the conversion; this is derived using WebPIMMS, which yields a stellar coronal flux ratio of 1.15 between the 0.1-2.4\,keV and the 0.2-2\,keV flux for a typical coronal temperature of 0.3\,keV. We then add the X-ray and EUV to obtain the XUV flux.

The resulting X-ray fluxes in the 0.2-2.0 keV band at the planetary orbits are shown in Fig.~\ref{fig:fxplanet}. The XUV fluxes then to be a factor of about five  to ten higher than the X-ray fluxes alone. We immediately see the X-ray luminosity distribution of cool stars depicted in the vertical spread at any given planetary orbital semi-major axis. The planets that are amenable to follow-up observations of currently ongoing mass loss are planets that are transiting and highly irradiated in the high-energy regime. We indicate the transiting planets by black circles and the new X-ray detections of host stars of transiting planets by additional black crosses. The clustering of transiting planets at close orbital distances comes from the geometric probability of a planet at a given inclination actually transiting, which strongly decreases for larger semi-major axes. Among the about 90 transiting planets with X-ray detected host stars, 26 stem from new eROSITA discoveries. Comparing the irradiation fluxes with the known evaporating exoplanets HD~189733~b and GJ~436~b (see discussion in section~\ref{discussion}), we find that a total of 50 exoplanets in our sample experience irradiation levels in excess of the one experienced by GJ~436~b, indicating that these exoplanets may be undergoing directly observable evaporation at the moment.

To determine the mass loss, data on both the planetary mass and radius is necessary. However, depending on the discovery method of a given planet, only the planetary mass or the radius may be known, but not necessarily both. In those cases we used the mass-radius relationship by \citet{Chen2017} to estimate a planet's radius from its mass, or its mass from its radius. \citet{Chen2017} divide planets into a ''terran'' regime with radii below $2R_\oplus$ where masses are dominated by the rocky core, and a ''Neptunian'' regime for planets with radii of $2R_\oplus$ or more where the gaseous envelope contains a significant fraction of the total mass. In our final mass loss estimates we only include numeric estimates for exoplanets with radii larger than $1.6R_\oplus$, based on \citet{Rogers2015} who showed that exoplanets smaller than $1.6R_\oplus$ are likely fully rocky and therefore unlikely to undergo any significant atmospheric mass loss.

We show the resulting mass loss estimates in Fig.~\ref{fig:mdotplanet}; a table of estimated mass loss rates and the related X-ray quantities is available as an electronic data table (see Appendix). The highest mass loss rates are found, as expected, for planets in very close orbits around their host stars. Here the vertical spread comes from both the X-ray luminosity distribution of the host stars and the spread in individual exoplanet masses and radii, which influences the estimated mass loss rates as well. 
We find four new eROSITA measurements 
with expected mass loss rates higher than for HD~189733~b, and 14 new eROSITA discoveries with expected mass loss rates higher than for the Neptune GJ~436~b.


\section{Discussion}\label{discussion}

\subsection{Exoplanet mass loss rates in context}

While exoplanet evaporation rates can be estimated under the energy-limited mass loss regime, there are also direct observations of ongoing evaporation for some planets. One possibility for direct mass loss observations is through transit observations in the ultra-violet Ly-$\alpha$ line of hydrogen. The observed line shape is quite complex, even in the absence of exoplanetary mass loss; the host star produces emission in Ly-$\alpha$, which is then partially absorbed by the interstellar medium, and finally, also Earth's geocorona may add to the detected photons at the Ly-$\alpha$ wavelength. An actively evaporating exoplanetary atmosphere can be detected in the wings of the line, where planetary hydrogen moving at moderately high velocities causes extra absorption in the blue or red wing (or both) during the planet's transit. This has first been successfully observed for the Hot Jupiter HD~209458~b \citep{Vidal-Madjar2003}, and subsequent observations have targeted other Hot Jupiters and even Neptunic planets. We focus here on the Hot Jupiter HD~189733~b and the mini-Neptune GJ~436~b for our comparisons to the estimated mass loss rates of other exoplanets.

HD~189733~b is a well-studied transiting Hot Jupiter, orbiting a K0 star at a distance of about 20\,pc from the Sun. The optical brightness of the host star makes this exoplanet one of the best-studied targets for transmission spectroscopy. Its absorption signature in the hydrogen Ly-$\alpha$ line showed that the planet is undergoing mass loss \citep{Lecavelier2010}. Since Ly-$\alpha$ observations can only quantify atomic (not ionised) hydrogen in the planetary atmosphere that is moving at relatively high speeds, determining mass loss rates from such observations typically requires some additional model assumptions. \citet{Lecavelier2010} estimate the current mass loss rate to be in the range of $10^{9}$ and $ 10^{11}$\,gs$^{-1}$. Other estimates for the mass loss have been made such as the model by \citet{Chadney2017}, who calculate an upper limit to the mass loss rate of HD 189733b of the order of $10^{7}$ and $10^{12} gs^{-1}$ during stellar flares with large scale proton events. 
Our calculation, based on the energy-limited escape model, estimates a current mass loss rate of about $1.8 \times 10^{11} gs^{-1}$, which is at the upper end, but compatible with the estimates based on hydrogen Ly-$\alpha$ observations.

In contrast, GJ~436~b is a mini-Neptune in a close orbit around a nearby (9.75\,pc) M dwarf that is only mildly X-ray active. This exoplanet has been observed to expel a spectacularly large hydrogen tail, as was first presented by \citet{Kulow2014} and then analysed in more detail by \citet{Ehrenreich2015}. The atomic hydrogen tail observed in the Ly-$\alpha$ line covers almost half of the stellar disk during transits, with the egress being delayed by several hours compared to the broadband optical transit. This is consistent with an extended tail-like structure consisting of the evaporating atmosphere. While the hydrogen transit signal is strong and well observable, the modelled mass loss rate is rather modest with about $4\times 10^{6}$ and $10^{9} gs^{-1}$ \citep{Kulow2014}. Simulations conducted by \cite{Villarreal2021} calculate a higher mass loss rate of up to $10^{10} gs^{-1}$. 
Our own results based on the high-energy irradiation estimates a current mass loss rate of about $1.0 \times 10^{9}$\,gs$^{-1}$, being bracketed by the hydrogen observations and simulations. This exoplanet shows that even more moderate mass loss rates can produce strong observable signatures.

\subsection{Interesting individual targets identified with eROSITA data}

Transiting exoplanets with a high level of XUV irradiation are suitable for follow-up observations of ongoing mass loss, for example through transmission spectroscopy in the ultraviolet hydrogen Ly-$\alpha$ line or the metastable lines of helium (He{\sc{I}}\,10830) in the infrared. We discuss some particularly interesting systems here briefly in terms of the planetary high-energy environment and mass loss. 


\textit{TOI-251}: TOI-251 is a solar-mass dwarf star with an apparent \textit{Gaia} brightness of $G=9.8$\,mag, located at a distance of 99.5\,pc from the Sun; it is a relatively young star with an age of 40--320\,Myr, estimated from its rotation and magnetic activity \citep{Zhou2021}. TOI-251 hosts a mini-Neptune with a radius of $2.7 R_\oplus$ that orbits its host star in 4.9\,days. We detect the host star with a high X-ray luminosity in the 0.2-2.0 keV band of $L_X = 1.5\times10^{29}$\,erg\,$s^{-1}$ in eRASS1 and $L_X = 1.9\times10^{29}$\,erg\,$s^{-1}$ in eRASS2. The source was previously detected in a pointed observation with ROSAT, with the reported X-ray luminosity being slightly lower with $L_X = 1.1\times10^{29}$\,erg\,$s^{-1}$ \citep{Zhou2021}. We derive that the orbiting mini-Neptune experiences an X-ray irradiation flux of about 16000\,erg\,s$^{-1}$\,cm$^{-2}$, corresponding to an estimated mass loss of about $3\times10^9$\,g\,s$^{-1}$. This is higher than the estimated mass loss rate of the known evaporating mini-Neptune GJ~436~b by a factor of ten, which makes this a highly interesting target for follow-up observations. Furthermore, this planet currently straddles the so-called ''evaporation gap'' in the radius distribution of planets \citep{Fulton2017}, indicating that it may currently undergo a marked change in radius due to evaporation. \citet{Poppenhaeger2021} recently showed for another young star-planet system, V1298~Tau and its four planets, that the early rotational evolution of the host star can make significant difference in the initial-to-final radius relationship of exoplanets. It is interesting in this context that TOI-251 seems to have already arrived on the ''slow/I-type'' sequence \citep{Barnes2010} of the stellar colour-rotation diagram \citep{Zhou2021}, making its future magnetic activity evolution and therefore the future planetary mass loss evolution relatively well predictable.

\textit{GJ 143}: GJ~143 is a bright K dwarf located at a distance of 16.3\,pc from the Sun, with an apparent \textit{Gaia} brightness of $G=7.7$\,mag. It hosts two planets, planet $b$ is a Neptunic planet in an orbit of 35.6\,days \citep{Trifonov2019} and planet $c$ is a small rocky planet in a closer orbit of 7.8\,days \citep{Dragomir2019}. The host star is only moderately X-ray bright with a luminosity of $L_X=1.7\times10^{27}$\,erg\,s$^{-1}$ in the 0.2-2.0 keV band. However, this is enough to create a high-energy environment of roughly similar intensity as the one that is present for the evaporating Neptune GJ~436~b, which experiences an X-ray irradiation flux of $F_\mathrm{X,\,pl}=50$\,erg\,s$^{-1}$\,$cm^{-2}$; planet $b$ and $c$ experience irradiation fluxes of $F_\mathrm{X,\,pl}=16$ and $123$\,erg\,s$^{-1}$\,$cm^{-2}$, respectively. While planet $c$ is now rocky and does likely not have a thick gaseous envelope that it can lose mass from, it is possible that it was formed with a hydrogen-helium envelope that it lost over time, especially during higher X-ray activity epochs in the youth of the host star. We estimate that the larger planet $b$ loses its atmosphere at a rate of about $5\times 10^7$\,g\,s$^{-1}$, i.e.\ about an order of magnitude less than GJ~436~b. However, given that the host star is optically brighter by a factor of about six, GJ~143~b's ongoing evaporation may well be observable with current instrumentation.

\textit{K2-198\,bcd}: K2-198 is a K dwarf star that is located at a distance of 110.6\,pc from the Sun, with an apparent \textit{Gaia} brightness of $G=11.0$\,mag. It hast three known transiting planets \citep{Mayo2018, Hedges2019}: the innermost planet c orbits at a period of 3.4 days and has a radius of 1.4\,$R_\oplus$ \citep{Hedges2019}, which places it in the regime of rocky planets. The middle planet d is a mini-Neptune with an orbital period of 7.5\,days and a radius of 2.4\,$R_\oplus$ \citep{Hedges2019}. The largest planet b is a Saturn-like planet in a wider orbit with an orbital period of 25.9 days and a radius of 4.2\,$R_\oplus$ \citep{Mayo2018}. The eROSITA data determined the host star's X-ray luminosity to be $7.9\times 10^{28}$\,erg\,s$^{-1}$ in the 0.2-2.0 keV band. This places all three planets in an intense X-ray irradiation regime with fluxes at the planetary orbits of about $F_{\mathrm{X,\,pl}}=17020/5890/1950$\,erg\,s$^{-1}$cm$^{-2}$ for planets $c$, $d$, and $b$, respectively. The middle planet $d$ is more strongly irradiated than the evaporating mini-Neptune GJ~436~b and can be expected to actively lose mass at a high estimated rate of $4\times 10^{10}$\,g\,s$^{-1}$. The innermost planet is likely to be rocky and might therefore not undergo any significantly mass loss any more. However, it is possible that it was formed with a primordial hydrogen-helium envelope which has been evaporated completely in the youth of the system. The heaviest planet $b$ experiences an intermediate intensity of high-energy irradiation, which is lower than for the known evaporating Hot Jupiter HD~189733~b, but higher than for GJ~436~b, and we estimate it to undergo mass loss at a rate of $2\times 10^{10}$\,g\,s$^{-1}$.

\textit{K2-240\,bc}: K2-240 is an early M dwarf which was discovered to host two transiting mini-Neptunes \citep{DiezAlonso2018}. It is located at a distance of 72.9\,pc from the Sun and an apparent \textit{Gaia} brightness of $G=12.6$\,mag; we determine its X-ray luminosity to be $3.7\times 10^{28}$\,erg\,s$^{-1}$ in the 0.2-2.0 keV band. The two mini-Neptunes orbit the star in 6.0 and 20.5\,days each. Their X-ray irradiation levels are well above those of the evaporating Neptune GJ~436\,b with $F_\mathrm{X,\,pl}= 5020$ and $980$\,erg\,s$^{-1}$\,cm$^{-2}$ for planets $b$ and $c$, respectively. We estimate their mass loss rates to be $\dot{M}= 3\times 10^{10}$ and $5\times10^{9}$\,g\,s$^{-1}$ for planets $b$ and $c$. This is higher than for GJ~436~b by more than an order of magnitude for planet $b$; these two planets may make good targets for follow-up observations of ongoing evaporation.

\section{Conclusion}\label{conclusion}

We have presented a catalogue of X-ray luminosities of exo\-pla\-net host stars, high-energy irradiation levels of exoplanets and their estimated atmospheric mass loss rates. We have combined new data from the eROSITA mission's first and second all-sky surveys (eRASS1 and eRASS2), and have amended our catalogue with archival data from ROSAT, \textit{XMM-Newton}, and \textit{Chandra}. We have presented 
high-energy irradiation levels for 329 exoplanets, 108 of them stemming from first-time detections with eROSITA, and mass loss estimates for 287 exoplanets, 96 of them derived from first-time eROSITA detections. Particularly interesting targets for follow-up observations of ongoing mass loss have been found, among them two multi-planet systems which can lead to unique insights on the evolution of exoplanetary atmospheres over time.

\begin{acknowledgements}
The authors thank Dr.\ Iris Traulsen for fruitful discussions of the \textit{XMM-Newton} source catalogs, and Laura Ketzer for helpful input on the mass-radius relationship of exoplanets. This work is based on data from eROSITA, the primary instrument aboard SRG, a joint Russian-German science mission supported by the Russian Space Agency (Roskosmos), in the interests of the Russian Academy of Sciences represented by its Space Research Institute (IKI), and the Deutsches Zentrum für Luft- und Raumfahrt (DLR). The SRG spacecraft was built by Lavochkin Association (NPOL) and its subcontractors, and is operated by NPOL with support from the Max Planck Institute for Extraterrestrial Physics (MPE).
The development and construction of the eROSITA X-ray instrument was led by MPE, with contributions from the Dr. Karl Remeis Observatory Bamberg \& ECAP (FAU Erlangen-Nuernberg), the University of Hamburg Observatory, the Leibniz Institute for Astrophysics Potsdam (AIP), and the Institute for Astronomy and Astrophysics of the University of Tübingen, with the support of DLR and the Max Planck Society. The Argelander Institute for Astronomy of the University of Bonn and the Ludwig Maximilians Universität Munich also participated in the science preparation for eROSITA. The eROSITA data shown here were processed using the eSASS software system developed by the German eROSITA consortium. This work has made use of data from the \textit{Chandra} X-ray Observatory, the ROSAT mission, and the \textit{XMM-Newton} mission.

\end{acknowledgements}

\bibliographystyle{aa}
\bibliography{biblio}

\begin{thebibliography}{83}
\expandafter\ifx\csname natexlab\endcsname\relax\def\natexlab#1{#1}\fi

\bibitem[{{Andrae} {et~al.}(2018){Andrae}, {Fouesneau}, {Creevey}, {Ordenovic},
  {Mary}, {Burlacu}, {Chaoul}, {Jean-Antoine-Piccolo}, {Kordopatis}, {Korn},
  {Lebreton}, {Panem}, {Pichon}, {Th{\'e}venin}, {Walmsley}, \&
  {Bailer-Jones}}]{Andrae2018}
{Andrae}, R., {Fouesneau}, M., {Creevey}, O., {et~al.} 2018, \aap, 616, A8

\bibitem[{{Arnaud}(1996)}]{Arnaud1996}
{Arnaud}, K.~A. 1996, in Astronomical Society of the Pacific Conference Series,
  Vol. 101, Astronomical Data Analysis Software and Systems V, ed. G.~H.
  {Jacoby} \& J.~{Barnes}, 17

\bibitem[{{Ayres} {et~al.}(2003){Ayres}, {Brown}, \& {Harper}}]{Ayres2003}
{Ayres}, T.~R., {Brown}, A., \& {Harper}, G.~M. 2003, \apj, 598, 610

\bibitem[{{Baraffe} {et~al.}(2010){Baraffe}, {Chabrier}, \&
  {Barman}}]{Baraffe2010}
{Baraffe}, I., {Chabrier}, G., \& {Barman}, T. 2010, Reports on Progress in
  Physics, 73, 016901

\bibitem[{{Barnes}(2010)}]{Barnes2010}
{Barnes}, S.~A. 2010, \apj, 722, 222

\bibitem[{{Boller} {et~al.}(2016){Boller}, {Freyberg}, {Tr{\"u}mper}, {Haberl},
  {Voges}, \& {Nandra}}]{Boller2016}
{Boller}, T., {Freyberg}, M.~J., {Tr{\"u}mper}, J., {et~al.} 2016, \aap, 588,
  A103

\bibitem[{{Bours} {et~al.}(2014){Bours}, {Marsh}, {Breedt}, {Copperwheat},
  {Dhillon}, {Leckngam}, {Littlefair}, {Parsons}, \& {Prasit}}]{Bours2014}
{Bours}, M.~C.~P., {Marsh}, T.~R., {Breedt}, E., {et~al.} 2014, \mnras, 445,
  1924

\bibitem[{{Brunner} {et~al.}(2021){Brunner}, {Liu}, {Lamer}, {Georgakakis},
  {Merloni}, \& {Brusa}}]{Brunner2021efeds}
{Brunner}, H., {Liu}, T., {Lamer}, G., {et~al.} 2021, \aap, xxx, xx

\bibitem[{{Chadney} {et~al.}(2015){Chadney}, {Galand}, {Unruh}, {Koskinen}, \&
  {Sanz-Forcada}}]{Chadney2015}
{Chadney}, J.~M., {Galand}, M., {Unruh}, Y.~C., {Koskinen}, T.~T., \&
  {Sanz-Forcada}, J. 2015, \icarus, 250, 357

\bibitem[{{Chadney} {et~al.}(2017){Chadney}, {Koskinen}, {Galand}, {Unruh}, \&
  {Sanz-Forcada}}]{Chadney2017}
{Chadney}, J.~M., {Koskinen}, T.~T., {Galand}, M., {Unruh}, Y.~C., \&
  {Sanz-Forcada}, J. 2017, \aap, 608, A75

\bibitem[{{Chen} \& {Kipping}(2017)}]{Chen2017}
{Chen}, J. \& {Kipping}, D. 2017, \apj, 834, 17

\bibitem[{{Coffaro} {et~al.}(2020){Coffaro}, {Stelzer}, {Orlando}, {Hall},
  {Metcalfe}, {Wolter}, {Mittag}, {Sanz-Forcada}, {Schneider}, \&
  {Ducci}}]{Coffaro2020}
{Coffaro}, M., {Stelzer}, B., {Orlando}, S., {et~al.} 2020, \aap, 636, A49

\bibitem[{{Cohen} {et~al.}(2015){Cohen}, {Ma}, {Drake}, {Glocer}, {Garraffo},
  {Bell}, \& {Gombosi}}]{Cohen2015}
{Cohen}, O., {Ma}, Y., {Drake}, J.~J., {et~al.} 2015, \apj, 806, 41

\bibitem[{{D{\'\i}ez Alonso} {et~al.}(2018){D{\'\i}ez Alonso}, {Gonz{\'a}lez
  Hern{\'a}ndez}, {Su{\'a}rez G{\'o}mez}, {Aguado}, {Gonz{\'a}lez
  Guti{\'e}rrez}, {Su{\'a}rez Mascare{\~n}o}, {Cabrera-Lavers},
  {Gonz{\'a}lez-Nuevo}, {Toledo-Padr{\'o}n}, {Gracia}, {de Cos Juez}, \&
  {Rebolo}}]{DiezAlonso2018}
{D{\'\i}ez Alonso}, E., {Gonz{\'a}lez Hern{\'a}ndez}, J.~I., {Su{\'a}rez
  G{\'o}mez}, S.~L., {et~al.} 2018, \mnras, 480, L1

\bibitem[{{Dong} {et~al.}(2017){Dong}, {Lingam}, {Ma}, \& {Cohen}}]{Dong2017}
{Dong}, C., {Lingam}, M., {Ma}, Y., \& {Cohen}, O. 2017, \apjl, 837, L26

\bibitem[{{Dragomir} {et~al.}(2019){Dragomir}, {Teske}, {G{\"u}nther},
  {S{\'e}gransan}, {Burt}, {Huang}, {Vanderburg}, {Matthews}, {Dumusque},
  {Stassun}, {Pepper}, {Ricker}, {Vanderspek}, {Latham}, {Seager}, {Winn},
  {Jenkins}, {Beatty}, {Bouchy}, {Brown}, {Butler}, {Ciardi}, {Crane},
  {Eastman}, {Fossati}, {Francis}, {Fulton}, {Gaudi}, {Goeke}, {James},
  {Klaus}, {Kuhn}, {Lovis}, {Lund}, {McDermott}, {Paegert}, {Pepe},
  {Rodriguez}, {Sha}, {Shectman}, {Shporer}, {Siverd}, {Garcia Soto},
  {Stevens}, {Twicken}, {Udry}, {Villanueva}, {Wang}, {Wohler}, {Yao}, \&
  {Zhan}}]{Dragomir2019}
{Dragomir}, D., {Teske}, J., {G{\"u}nther}, M.~N., {et~al.} 2019, \apjl, 875,
  L7

\bibitem[{{Ehrenreich} {et~al.}(2015){Ehrenreich}, {Bourrier}, {Wheatley},
  {Lecavelier des Etangs}, {H{\'e}brard}, {Udry}, {Bonfils}, {Delfosse},
  {D{\'e}sert}, {Sing}, \& {Vidal-Madjar}}]{Ehrenreich2015}
{Ehrenreich}, D., {Bourrier}, V., {Wheatley}, P.~J., {et~al.} 2015, \nat, 522,
  459

\bibitem[{{Evans} \& {Civano}(2018)}]{Evans2018}
{Evans}, I.~N. \& {Civano}, F. 2018, Astronomy and Geophysics, 59, 2.17

\bibitem[{{Evans} {et~al.}(2010){Evans}, {Primini}, {Glotfelty}, {Anderson},
  {Bonaventura}, {Chen}, {Davis}, {Doe}, {Evans}, {Fabbiano}, {Galle}, {Gibbs},
  {Grier}, {Hain}, {Hall}, {Harbo}, {He}, {Houck}, {Karovska}, {Kashyap},
  {Lauer}, {McCollough}, {McDowell}, {Miller}, {Mitschang}, {Morgan},
  {Mossman}, {Nichols}, {Nowak}, {Plummer}, {Refsdal}, {Rots}, {Siemiginowska},
  {Sundheim}, {Tibbetts}, {Van Stone}, {Winkelman}, \& {Zografou}}]{Evans2010}
{Evans}, I.~N., {Primini}, F.~A., {Glotfelty}, K.~J., {et~al.} 2010, \apjs,
  189, 37

\bibitem[{{Fortney} \& {Nettelmann}(2010)}]{Fortney2010}
{Fortney}, J.~J. \& {Nettelmann}, N. 2010, \ssr, 152, 423

\bibitem[{{France} {et~al.}(2013){France}, {Froning}, {Linsky}, {Roberge},
  {Stocke}, {Tian}, {Bushinsky}, {D{\'e}sert}, {Mauas}, {Vieytes}, \&
  {Walkowicz}}]{France2013}
{France}, K., {Froning}, C.~S., {Linsky}, J.~L., {et~al.} 2013, \apj, 763, 149

\bibitem[{{Fulton} {et~al.}(2017){Fulton}, {Petigura}, {Howard}, {Isaacson},
  {Marcy}, {Cargile}, {Hebb}, {Weiss}, {Johnson}, {Morton}, {Sinukoff},
  {Crossfield}, \& {Hirsch}}]{Fulton2017}
{Fulton}, B.~J., {Petigura}, E.~A., {Howard}, A.~W., {et~al.} 2017, \aj, 154,
  109

\bibitem[{{Gaia Collaboration} {et~al.}(2018){Gaia Collaboration}, {Brown},
  {Vallenari}, {Prusti}, {de Bruijne}, {Babusiaux}, {Bailer-Jones}, {Biermann},
  {Evans}, {Eyer}, {Jansen}, {Jordi}, {Klioner}, {Lammers}, {Lindegren},
  {Luri}, {Mignard}, {Panem}, {Pourbaix}, {Randich}, {Sartoretti}, {Siddiqui},
  {Soubiran}, {van Leeuwen}, {Walton}, {Arenou}, {Bastian}, {Cropper},
  {Drimmel}, {Katz}, {Lattanzi}, {Bakker}, {Cacciari}, {Casta{\~n}eda},
  {Chaoul}, {Cheek}, {De Angeli}, {Fabricius}, {Guerra}, {Holl}, {Masana},
  {Messineo}, {Mowlavi}, {Nienartowicz}, {Panuzzo}, {Portell}, {Riello},
  {Seabroke}, {Tanga}, {Th{\'e}venin}, {Gracia-Abril}, {Comoretto},
  {Garcia-Reinaldos}, {Teyssier}, {Altmann}, {Andrae}, {Audard},
  {Bellas-Velidis}, {Benson}, {Berthier}, {Blomme}, {Burgess}, {Busso},
  {Carry}, {Cellino}, {Clementini}, {Clotet}, {Creevey}, {Davidson}, {De
  Ridder}, {Delchambre}, {Dell'Oro}, {Ducourant},
  {Fern{\'a}ndez-Hern{\'a}ndez}, {Fouesneau}, {Fr{\'e}mat}, {Galluccio},
  {Garc{\'\i}a-Torres}, {Gonz{\'a}lez-N{\'u}{\~n}ez}, {Gonz{\'a}lez-Vidal},
  {Gosset}, {Guy}, {Halbwachs}, {Hambly}, {Harrison}, {Hern{\'a}ndez},
  {Hestroffer}, {Hodgkin}, {Hutton}, {Jasniewicz}, {Jean-Antoine-Piccolo},
  {Jordan}, {Korn}, {Krone-Martins}, {Lanzafame}, {Lebzelter}, {L{\"o}ffler},
  {Manteiga}, {Marrese}, {Mart{\'\i}n-Fleitas}, {Moitinho}, {Mora}, {Muinonen},
  {Osinde}, {Pancino}, {Pauwels}, {Petit}, {Recio-Blanco}, {Richards},
  {Rimoldini}, {Robin}, {Sarro}, {Siopis}, {Smith}, {Sozzetti}, {S{\"u}veges},
  {Torra}, {van Reeven}, {Abbas}, {Abreu Aramburu}, {Accart}, {Aerts},
  {Altavilla}, {{\'A}lvarez}, {Alvarez}, {Alves}, {Anderson}, {Andrei},
  {Anglada Varela}, {Antiche}, {Antoja}, {Arcay}, {Astraatmadja}, {Bach},
  {Baker}, {Balaguer-N{\'u}{\~n}ez}, {Balm}, {Barache}, {Barata}, {Barbato},
  {Barblan}, {Barklem}, {Barrado}, {Barros}, {Barstow}, {Bartholom{\'e}
  Mu{\~n}oz}, {Bassilana}, {Becciani}, {Bellazzini}, {Berihuete}, {Bertone},
  {Bianchi}, {Bienaym{\'e}}, {Blanco-Cuaresma}, {Boch}, {Boeche}, {Bombrun},
  {Borrachero}, {Bossini}, {Bouquillon}, {Bourda}, {Bragaglia}, {Bramante},
  {Breddels}, {Bressan}, {Brouillet}, {Br{\"u}semeister}, {Brugaletta},
  {Bucciarelli}, {Burlacu}, {Busonero}, {Butkevich}, {Buzzi}, {Caffau},
  {Cancelliere}, {Cannizzaro}, {Cantat-Gaudin}, {Carballo}, {Carlucci},
  {Carrasco}, {Casamiquela}, {Castellani}, {Castro-Ginard}, {Charlot},
  {Chemin}, {Chiavassa}, {Cocozza}, {Costigan}, {Cowell}, {Crifo}, {Crosta},
  {Crowley}, {Cuypers}, {Dafonte}, {Damerdji}, {Dapergolas}, {David}, {David},
  {de Laverny}, {De Luise}, {De March}, {de Martino}, {de Souza}, {de Torres},
  {Debosscher}, {del Pozo}, {Delbo}, {Delgado}, {Delgado}, {Di Matteo},
  {Diakite}, {Diener}, {Distefano}, {Dolding}, {Drazinos}, {Dur{\'a}n},
  {Edvardsson}, {Enke}, {Eriksson}, {Esquej}, {Eynard Bontemps}, {Fabre},
  {Fabrizio}, {Faigler}, {Falc{\~a}o}, {Farr{\`a}s Casas}, {Federici},
  {Fedorets}, {Fernique}, {Figueras}, {Filippi}, {Findeisen}, {Fonti},
  {Fraile}, {Fraser}, {Fr{\'e}zouls}, {Gai}, {Galleti}, {Garabato},
  {Garc{\'\i}a-Sedano}, {Garofalo}, {Garralda}, {Gavel}, {Gavras}, {Gerssen},
  {Geyer}, {Giacobbe}, {Gilmore}, {Girona}, {Giuffrida}, {Glass}, {Gomes},
  {Granvik}, {Gueguen}, {Guerrier}, {Guiraud}, {Guti{\'e}rrez-S{\'a}nchez},
  {Haigron}, {Hatzidimitriou}, {Hauser}, {Haywood}, {Heiter}, {Helmi}, {Heu},
  {Hilger}, {Hobbs}, {Hofmann}, {Holland}, {Huckle}, {Hypki}, {Icardi},
  {Jan{\ss}en}, {Jevardat de Fombelle}, {Jonker}, {Juh{\'a}sz}, {Julbe},
  {Karampelas}, {Kewley}, {Klar}, {Kochoska}, {Kohley}, {Kolenberg},
  {Kontizas}, {Kontizas}, {Koposov}, {Kordopatis}, {Kostrzewa-Rutkowska},
  {Koubsky}, {Lambert}, {Lanza}, {Lasne}, {Lavigne}, {Le Fustec}, {Le
  Poncin-Lafitte}, {Lebreton}, {Leccia}, {Leclerc}, {Lecoeur-Taibi},
  {Lenhardt}, {Leroux}, {Liao}, {Licata}, {Lindstr{\o}m}, {Lister}, {Livanou},
  {Lobel}, {L{\'o}pez}, {Managau}, {Mann}, {Mantelet}, {Marchal}, {Marchant},
  {Marconi}, {Marinoni}, {Marschalk{\'o}}, {Marshall}, {Martino}, {Marton},
  {Mary}, {Massari}, {Matijevi{\v{c}}}, {Mazeh}, {McMillan}, {Messina},
  {Michalik}, {Millar}, {Molina}, {Molinaro}, {Moln{\'a}r}, {Montegriffo},
  {Mor}, {Morbidelli}, {Morel}, {Morris}, {Mulone}, {Muraveva}, {Musella},
  {Nelemans}, {Nicastro}, {Noval}, {O'Mullane}, {Ord{\'e}novic},
  {Ord{\'o}{\~n}ez-Blanco}, {Osborne}, {Pagani}, {Pagano}, {Pailler},
  {Palacin}, {Palaversa}, {Panahi}, {Pawlak}, {Piersimoni}, {Pineau}, {Plachy},
  {Plum}, {Poggio}, {Poujoulet}, {Pr{\v{s}}a}, {Pulone}, {Racero}, {Ragaini},
  {Rambaux}, {Ramos-Lerate}, {Regibo}, {Reyl{\'e}}, {Riclet}, {Ripepi}, {Riva},
  {Rivard}, {Rixon}, {Roegiers}, {Roelens}, {Romero-G{\'o}mez}, {Rowell},
  {Royer}, {Ruiz-Dern}, {Sadowski}, {Sagrist{\`a} Sell{\'e}s}, {Sahlmann},
  {Salgado}, {Salguero}, {Sanna}, {Santana-Ros}, {Sarasso}, {Savietto},
  {Schultheis}, {Sciacca}, {Segol}, {Segovia}, {S{\'e}gransan}, {Shih},
  {Siltala}, {Silva}, {Smart}, {Smith}, {Solano}, {Solitro}, {Sordo}, {Soria
  Nieto}, {Souchay}, {Spagna}, {Spoto}, {Stampa}, {Steele},
  {Steidelm{\"u}ller}, {Stephenson}, {Stoev}, {Suess}, {Surdej}, {Szabados},
  {Szegedi-Elek}, {Tapiador}, {Taris}, {Tauran}, {Taylor}, {Teixeira},
  {Terrett}, {Teyssandier}, {Thuillot}, {Titarenko}, {Torra Clotet}, {Turon},
  {Ulla}, {Utrilla}, {Uzzi}, {Vaillant}, {Valentini}, {Valette}, {van Elteren},
  {Van Hemelryck}, {van Leeuwen}, {Vaschetto}, {Vecchiato}, {Veljanoski},
  {Viala}, {Vicente}, {Vogt}, {von Essen}, {Voss}, {Votruba}, {Voutsinas},
  {Walmsley}, {Weiler}, {Wertz}, {Wevers}, {Wyrzykowski}, {Yoldas},
  {{\v{Z}}erjal}, {Ziaeepour}, {Zorec}, {Zschocke}, {Zucker}, {Zurbach}, \&
  {Zwitter}}]{gaiadr2}
{Gaia Collaboration}, {Brown}, A.~G.~A., {Vallenari}, A., {et~al.} 2018, \aap,
  616, A1

\bibitem[{{Garmire} {et~al.}(2003){Garmire}, {Bautz}, {Ford}, {Nousek}, \&
  {Ricker}}]{Garmire2003}
{Garmire}, G.~P., {Bautz}, M.~W., {Ford}, P.~G., {Nousek}, J.~A., \& {Ricker},
  George~R., J. 2003, in Society of Photo-Optical Instrumentation Engineers
  (SPIE) Conference Series, Vol. 4851, X-Ray and Gamma-Ray Telescopes and
  Instruments for Astronomy., ed. J.~E. {Truemper} \& H.~D. {Tananbaum}, 28--44

\bibitem[{{Go{\'z}dziewski} {et~al.}(2015){Go{\'z}dziewski}, {S{\l}owikowska},
  {Dimitrov}, {Krzeszowski}, {{\.Z}ejmo}, {Kanbach}, {Burwitz}, {Rau},
  {Irawati}, {Richichi}, {Gawro{\'n}ski}, {Nowak}, {Nasiroglu}, \&
  {Kubicki}}]{Gozdziewski2015}
{Go{\'z}dziewski}, K., {S{\l}owikowska}, A., {Dimitrov}, D., {et~al.} 2015,
  \mnras, 448, 1118

\bibitem[{{G{\"u}del}(2004)}]{Guedel2004}
{G{\"u}del}, M. 2004, \aapr, 12, 71

\bibitem[{{G{\"u}nther} {et~al.}(2012){G{\"u}nther}, {Wolk}, {Drake}, {Lisse},
  {Robrade}, \& {Schmitt}}]{Guenther2012}
{G{\"u}nther}, H.~M., {Wolk}, S.~J., {Drake}, J.~J., {et~al.} 2012, \apj, 750,
  78

\bibitem[{{Hedges} {et~al.}(2019){Hedges}, {Saunders}, {Barentsen}, {Coughlin},
  {Cardoso}, {Kostov}, {Dotson}, \& {Cody}}]{Hedges2019}
{Hedges}, C., {Saunders}, N., {Barentsen}, G., {et~al.} 2019, \apjl, 880, L5

\bibitem[{{Hempel} {et~al.}(2005){Hempel}, {Robrade}, {Ness}, \&
  {Schmitt}}]{Hempel2005}
{Hempel}, M., {Robrade}, J., {Ness}, J.~U., \& {Schmitt}, J.~H.~M.~M. 2005,
  \aap, 440, 727

\bibitem[{{Jansen} {et~al.}(2001){Jansen}, {Lumb}, {Altieri}, {Clavel}, {Ehle},
  {Erd}, {Gabriel}, {Guainazzi}, {Gondoin}, {Much}, {Munoz}, {Santos},
  {Schartel}, {Texier}, \& {Vacanti}}]{Jansen2001}
{Jansen}, F., {Lumb}, D., {Altieri}, B., {et~al.} 2001, \aap, 365, L1

\bibitem[{{Johnstone} {et~al.}(2020){Johnstone}, {Bartel}, \&
  {G{\"u}del}}]{Johnstone2020arXiv}
{Johnstone}, C.~P., {Bartel}, M., \& {G{\"u}del}, M. 2020, arXiv e-prints,
  arXiv:2009.07695

\bibitem[{{King} \& {Wheatley}(2021)}]{King2021}
{King}, G.~W. \& {Wheatley}, P.~J. 2021, \mnras, 501, L28

\bibitem[{{Kulow} {et~al.}(2014){Kulow}, {France}, {Linsky}, \&
  {Loyd}}]{Kulow2014}
{Kulow}, J.~R., {France}, K., {Linsky}, J., \& {Loyd}, R.~O.~P. 2014, \apj,
  786, 132

\bibitem[{{Lecavelier Des Etangs} {et~al.}(2010){Lecavelier Des Etangs},
  {Ehrenreich}, {Vidal-Madjar}, {Ballester}, {D{\'e}sert}, {Ferlet},
  {H{\'e}brard}, {Sing}, {Tchakoumegni}, \& {Udry}}]{Lecavelier2010}
{Lecavelier Des Etangs}, A., {Ehrenreich}, D., {Vidal-Madjar}, A., {et~al.}
  2010, \aap, 514, A72

\bibitem[{{Lindegren} {et~al.}(2018){Lindegren}, {Hern{\'a}ndez}, {Bombrun},
  {Klioner}, {Bastian}, {Ramos-Lerate}, {de Torres}, {Steidelm{\"u}ller},
  {Stephenson}, {Hobbs}, {Lammers}, {Biermann}, {Geyer}, {Hilger}, {Michalik},
  {Stampa}, {McMillan}, {Casta{\~n}eda}, {Clotet}, {Comoretto}, {Davidson},
  {Fabricius}, {Gracia}, {Hambly}, {Hutton}, {Mora}, {Portell}, {van Leeuwen},
  {Abbas}, {Abreu}, {Altmann}, {Andrei}, {Anglada}, {Balaguer-N{\'u}{\~n}ez},
  {Barache}, {Becciani}, {Bertone}, {Bianchi}, {Bouquillon}, {Bourda},
  {Br{\"u}semeister}, {Bucciarelli}, {Busonero}, {Buzzi}, {Cancelliere},
  {Carlucci}, {Charlot}, {Cheek}, {Crosta}, {Crowley}, {de Bruijne}, {de
  Felice}, {Drimmel}, {Esquej}, {Fienga}, {Fraile}, {Gai}, {Garralda},
  {Gonz{\'a}lez-Vidal}, {Guerra}, {Hauser}, {Hofmann}, {Holl}, {Jordan},
  {Lattanzi}, {Lenhardt}, {Liao}, {Licata}, {Lister}, {L{\"o}ffler},
  {Marchant}, {Martin-Fleitas}, {Messineo}, {Mignard}, {Morbidelli}, {Poggio},
  {Riva}, {Rowell}, {Salguero}, {Sarasso}, {Sciacca}, {Siddiqui}, {Smart},
  {Spagna}, {Steele}, {Taris}, {Torra}, {van Elteren}, {van Reeven}, \&
  {Vecchiato}}]{Lindegren2018}
{Lindegren}, L., {Hern{\'a}ndez}, J., {Bombrun}, A., {et~al.} 2018, \aap, 616,
  A2

\bibitem[{{Linsky} {et~al.}(2014){Linsky}, {Fontenla}, \&
  {France}}]{Linsky2014}
{Linsky}, J.~L., {Fontenla}, J., \& {France}, K. 2014, \apj, 780, 61

\bibitem[{{Lopez} {et~al.}(2012){Lopez}, {Fortney}, \& {Miller}}]{Lopez2012}
{Lopez}, E.~D., {Fortney}, J.~J., \& {Miller}, N. 2012, \apj, 761, 59

\bibitem[{{Loyd} {et~al.}(2018){Loyd}, {France}, {Youngblood}, {Schneider},
  {Brown}, {Hu}, {Segura}, {Linsky}, {Redfield}, {Tian}, {Rugheimer}, {Miguel},
  \& {Froning}}]{Loyd2018}
{Loyd}, R.~O.~P., {France}, K., {Youngblood}, A., {et~al.} 2018, \apj, 867, 71

\bibitem[{{Mayo} {et~al.}(2018){Mayo}, {Vanderburg}, {Latham}, {Bieryla},
  {Morton}, {Buchhave}, {Dressing}, {Beichman}, {Berlind}, {Calkins}, {Ciardi},
  {Crossfield}, {Esquerdo}, {Everett}, {Gonzales}, {Hirsch}, {Horch}, {Howard},
  {Howell}, {Livingston}, {Patel}, {Petigura}, {Schlieder}, {Scott}, {Schumer},
  {Sinukoff}, {Teske}, \& {Winters}}]{Mayo2018}
{Mayo}, A.~W., {Vanderburg}, A., {Latham}, D.~W., {et~al.} 2018, \aj, 155, 136

\bibitem[{{Mayor} \& {Queloz}(1995)}]{Mayor1995}
{Mayor}, M. \& {Queloz}, D. 1995, \nat, 378, 355

\bibitem[{{Merloni} {et~al.}(2012){Merloni}, {Predehl}, {Becker},
  {B{\"o}hringer}, {Boller}, {Brunner}, {Brusa}, {Dennerl}, {Freyberg},
  {Friedrich}, {Georgakakis}, {Haberl}, {Hasinger}, {Meidinger}, {Mohr},
  {Nandra}, {Rau}, {Reiprich}, {Robrade}, {Salvato}, {Santangelo}, {Sasaki},
  {Schwope}, {Wilms}, \& {German eROSITA Consortium}}]{Merloni2012}
{Merloni}, A., {Predehl}, P., {Becker}, W., {et~al.} 2012, arXiv e-prints,
  arXiv:1209.3114

\bibitem[{{Mugrauer}(2019)}]{Mugrauer2019}
{Mugrauer}, M. 2019, \mnras, 490, 5088

\bibitem[{{Murray} {et~al.}(1997){Murray}, {Chappell}, {Kenter}, {Kobayashi},
  {Kraft}, {Meehan}, {Zombeck}, {Fraser}, {Pearson}, {Lees}, {Brunton},
  {Pearce}, {Barbera}, {Collura}, \& {Serio}}]{Murray1997}
{Murray}, S.~S., {Chappell}, J.~H., {Kenter}, A.~T., {et~al.} 1997, in Society
  of Photo-Optical Instrumentation Engineers (SPIE) Conference Series, Vol.
  3114, EUV, X-Ray, and Gamma-Ray Instrumentation for Astronomy VIII, ed. O.~H.
  {Siegmund} \& M.~A. {Gummin}, 11--25

\bibitem[{{Murray-Clay} {et~al.}(2009){Murray-Clay}, {Chiang}, \&
  {Murray}}]{Murray-Clay2009}
{Murray-Clay}, R.~A., {Chiang}, E.~I., \& {Murray}, N. 2009, \apj, 693, 23

\bibitem[{{Newton} {et~al.}(2019){Newton}, {Mann}, {Tofflemire}, {Pearce},
  {Rizzuto}, {Vanderburg}, {Martinez}, {Wang}, {Ruffio}, {Kraus}, {Johnson},
  {Thao}, {Wood}, {Rampalli}, {Nielsen}, {Collins}, {Dragomir}, {Hellier},
  {Anderson}, {Barclay}, {Brown}, {Feiden}, {Hart}, {Isopi}, {Kielkopf},
  {Mallia}, {Nelson}, {Rodriguez}, {Stockdale}, {Waite}, {Wright}, {Lissauer},
  {Ricker}, {Vanderspek}, {Latham}, {Seager}, {Winn}, {Jenkins}, {Bouma},
  {Burke}, {Davies}, {Fausnaugh}, {Li}, {Morris}, {Mukai}, {Villase{\~n}or},
  {Villeneuva}, {De Rosa}, {Macintosh}, {Mengel}, {Okumura}, \&
  {Wittenmyer}}]{Newton2019}
{Newton}, E.~R., {Mann}, A.~W., {Tofflemire}, B.~M., {et~al.} 2019, \apjl, 880,
  L17

\bibitem[{{Nortmann} {et~al.}(2018){Nortmann}, {Pall{\'e}}, {Salz},
  {Sanz-Forcada}, {Nagel}, {Alonso-Floriano}, {Czesla}, {Yan}, {Chen},
  {Snellen}, {Zechmeister}, {Schmitt}, {L{\'o}pez-Puertas}, {Casasayas-Barris},
  {Bauer}, {Amado}, {Caballero}, {Dreizler}, {Henning}, {Lamp{\'o}n}, {Montes},
  {Molaverdikhani}, {Quirrenbach}, {Reiners}, {Ribas}, {S{\'a}nchez-L{\'o}pez},
  {Schneider}, \& {Zapatero Osorio}}]{Nortmann2018}
{Nortmann}, L., {Pall{\'e}}, E., {Salz}, M., {et~al.} 2018, Science, 362, 1388

\bibitem[{{Owen} \& {Adams}(2014)}]{Owen2014}
{Owen}, J.~E. \& {Adams}, F.~C. 2014, \mnras, 444, 3761

\bibitem[{{Owen} \& {Jackson}(2012)}]{Owen2012}
{Owen}, J.~E. \& {Jackson}, A.~P. 2012, \mnras, 425, 2931

\bibitem[{{Poppenhaeger} {et~al.}(2021){Poppenhaeger}, {Ketzer}, \&
  {Mallonn}}]{Poppenhaeger2021}
{Poppenhaeger}, K., {Ketzer}, L., \& {Mallonn}, M. 2021, \mnras, 500, 4560

\bibitem[{{Poppenhaeger} {et~al.}(2010){Poppenhaeger}, {Robrade}, \&
  {Schmitt}}]{Poppenhaeger2010}
{Poppenhaeger}, K., {Robrade}, J., \& {Schmitt}, J.~H.~M.~M. 2010, \aap, 515,
  A98

\bibitem[{{Poppenhaeger} {et~al.}(2013){Poppenhaeger}, {Schmitt}, \&
  {Wolk}}]{Poppenhaeger2013}
{Poppenhaeger}, K., {Schmitt}, J.~H.~M.~M., \& {Wolk}, S.~J. 2013, \apj, 773,
  62

\bibitem[{{Poppenhaeger} \& {Wolk}(2014)}]{Poppenhaeger2014}
{Poppenhaeger}, K. \& {Wolk}, S.~J. 2014, \aap, 565, L1

\bibitem[{{Predehl} {et~al.}(2021){Predehl}, {Andritschke}, {Arefiev},
  {Babyshkin}, {Batanov}, {Becker}, {B{\"o}hringer}, {Bogomolov}, {Boller},
  {Borm}, {Bornemann}, {Br{\"a}uninger}, {Br{\"u}ggen}, {Brunner}, {Brusa},
  {Bulbul}, {Buntov}, {Burwitz}, {Burkert}, {Clerc}, {Churazov}, {Coutinho},
  {Dauser}, {Dennerl}, {Doroshenko}, {Eder}, {Emberger}, {Eraerds},
  {Finoguenov}, {Freyberg}, {Friedrich}, {Friedrich}, {F{\"u}rmetz},
  {Georgakakis}, {Gilfanov}, {Granato}, {Grossberger}, {Gueguen}, {Gureev},
  {Haberl}, {H{\"a}lker}, {Hartner}, {Hasinger}, {Huber}, {Ji}, {Kienlin},
  {Kink}, {Korotkov}, {Kreykenbohm}, {Lamer}, {Lomakin}, {Lapshov}, {Liu},
  {Maitra}, {Meidinger}, {Menz}, {Merloni}, {Mernik}, {Mican}, {Mohr},
  {M{\"u}ller}, {Nandra}, {Nazarov}, {Pacaud}, {Pavlinsky}, {Perinati},
  {Pfeffermann}, {Pietschner}, {Ramos-Ceja}, {Rau}, {Reiffers}, {Reiprich},
  {Robrade}, {Salvato}, {Sanders}, {Santangelo}, {Sasaki}, {Scheuerle},
  {Schmid}, {Schmitt}, {Schwope}, {Shirshakov}, {Steinmetz}, {Stewart},
  {Str{\"u}der}, {Sunyaev}, {Tenzer}, {Tiedemann}, {Tr{\"u}mper}, {Voron},
  {Weber}, {Wilms}, \& {Yaroshenko}}]{Predehl2021}
{Predehl}, P., {Andritschke}, R., {Arefiev}, V., {et~al.} 2021, \aap, 647, A1

\bibitem[{{Raghavan} {et~al.}(2010){Raghavan}, {McAlister}, {Henry}, {Latham},
  {Marcy}, {Mason}, {Gies}, {White}, \& {ten Brummelaar}}]{Raghavan2010}
{Raghavan}, D., {McAlister}, H.~A., {Henry}, T.~J., {et~al.} 2010, \apjs, 190,
  1

\bibitem[{{Rogers}(2015)}]{Rogers2015}
{Rogers}, L.~A. 2015, \apj, 801, 41

\bibitem[{{Rosen} {et~al.}(2016){Rosen}, {Webb}, {Watson}, {Ballet}, {Barret},
  {Braito}, {Carrera}, {Ceballos}, {Coriat}, {Della Ceca}, {Denkinson},
  {Esquej}, {Farrell}, {Freyberg}, {Gris{\'e}}, {Guillout}, {Heil},
  {Koliopanos}, {Law-Green}, {Lamer}, {Lin}, {Martino}, {Michel}, {Motch},
  {Nebot Gomez-Moran}, {Page}, {Page}, {Page}, {Pakull}, {Pye}, {Read},
  {Rodriguez}, {Sakano}, {Saxton}, {Schwope}, {Scott}, {Sturm}, {Traulsen},
  {Yershov}, \& {Zolotukhin}}]{Rosen2016}
{Rosen}, S.~R., {Webb}, N.~A., {Watson}, M.~G., {et~al.} 2016, \aap, 590, A1

\bibitem[{{Salvato} {et~al.}(2018){Salvato}, {Buchner}, {Budav{\'a}ri},
  {Dwelly}, {Merloni}, {Brusa}, {Rau}, {Fotopoulou}, \& {Nandra}}]{Salvato2018}
{Salvato}, M., {Buchner}, J., {Budav{\'a}ri}, T., {et~al.} 2018, \mnras, 473,
  4937

\bibitem[{{Salz} {et~al.}(2015){Salz}, {Schneider}, {Czesla}, \&
  {Schmitt}}]{Salz2015massloss}
{Salz}, M., {Schneider}, P.~C., {Czesla}, S., \& {Schmitt}, J.~H.~M.~M. 2015,
  \aap, 576, A42

\bibitem[{{Salz} {et~al.}(2019){Salz}, {Schneider}, {Fossati}, {Czesla},
  {France}, \& {Schmitt}}]{Salz2019}
{Salz}, M., {Schneider}, P.~C., {Fossati}, L., {et~al.} 2019, \aap, 623, A57

\bibitem[{{Sanz-Forcada} {et~al.}(2011){Sanz-Forcada}, {Micela}, {Ribas},
  {Pollock}, {Eiroa}, {Velasco}, {Solano}, \&
  {Garc{\'\i}a-{\'A}lvarez}}]{Sanz-Forcada2011}
{Sanz-Forcada}, J., {Micela}, G., {Ribas}, I., {et~al.} 2011, \aap, 532, A6

\bibitem[{{Saxton} {et~al.}(2008){Saxton}, {Read}, {Esquej}, {Freyberg},
  {Altieri}, \& {Bermejo}}]{Saxton2008}
{Saxton}, R.~D., {Read}, A.~M., {Esquej}, P., {et~al.} 2008, \aap, 480, 611

\bibitem[{{Schmitt} {et~al.}(1995){Schmitt}, {Fleming}, \&
  {Giampapa}}]{Schmitt1995}
{Schmitt}, J. H.~M.~M., {Fleming}, T.~A., \& {Giampapa}, M.~S. 1995, \apj, 450,
  392

\bibitem[{{Schr{\"o}ter} {et~al.}(2011){Schr{\"o}ter}, {Czesla}, {Wolter},
  {M{\"u}ller}, {Huber}, \& {Schmitt}}]{Schroeter2011}
{Schr{\"o}ter}, S., {Czesla}, S., {Wolter}, U., {et~al.} 2011, \aap, 532, A3

\bibitem[{{Schwope} \& {Thinius}(2014)}]{Schwope2014}
{Schwope}, A.~D. \& {Thinius}, B.~D. 2014, Astronomische Nachrichten, 335, 357

\bibitem[{{Skinner} \& {G{\"u}del}(2020)}]{Skinner2020}
{Skinner}, S.~L. \& {G{\"u}del}, M. 2020, \apj, 888, 15

\bibitem[{{Spake} {et~al.}(2018){Spake}, {Sing}, {Evans}, {Oklop{\v{c}}i{\'c}},
  {}, {Bourrier}, {Kreidberg}, {Rackham}, {Irwin}, {Ehrenreich}, {Wyttenbach},
  {Wakeford}, {Zhou}, {Chubb}, {Nikolov}, {Goyal}, {Henry}, {Williamson},
  {Blumenthal}, {Anderson}, {Hellier}, {Charbonneau}, {Udry}, \&
  {Madhusudhan}}]{Spake2018}
{Spake}, J.~J., {Sing}, D.~K., {Evans}, T.~M., {et~al.} 2018, \nat, 557, 68

\bibitem[{{Str{\"u}der} {et~al.}(2001){Str{\"u}der}, {Briel}, {Dennerl},
  {Hartmann}, {Kendziorra}, {Meidinger}, {Pfeffermann}, {Reppin}, {Aschenbach},
  {Bornemann}, {Br{\"a}uninger}, {Burkert}, {Elender}, {Freyberg}, {Haberl},
  {Hartner}, {Heuschmann}, {Hippmann}, {Kastelic}, {Kemmer}, {Kettenring},
  {Kink}, {Krause}, {M{\"u}ller}, {Oppitz}, {Pietsch}, {Popp}, {Predehl},
  {Read}, {Stephan}, {St{\"o}tter}, {Tr{\"u}mper}, {Holl}, {Kemmer}, {Soltau},
  {St{\"o}tter}, {Weber}, {Weichert}, {von Zanthier}, {Carathanassis}, {Lutz},
  {Richter}, {Solc}, {B{\"o}ttcher}, {Kuster}, {Staubert}, {Abbey}, {Holland},
  {Turner}, {Balasini}, {Bignami}, {La Palombara}, {Villa}, {Buttler},
  {Gianini}, {Lain{\'e}}, {Lumb}, \& {Dhez}}]{Strueder2001}
{Str{\"u}der}, L., {Briel}, U., {Dennerl}, K., {et~al.} 2001, \aap, 365, L18

\bibitem[{{Sunyaev} {et~al.}(2021){Sunyaev}, {Arefiev}, {Babyshkin},
  {Bogomolov}, {Borisov}, {Buntov}, {Brunner}, {Burenin}, {Churazov},
  {Coutinho}, {Eder}, {Eismont}, {Freyberg}, {Gilfanov}, {Gureyev}, {Hasinger},
  {Khabibullin}, {Kolmykov}, {Komovkin}, {Krivonos}, {Lapshov}, {Levin},
  {Lomakin}, {Lutovinov}, {Medvedev}, {Merloni}, {Mernik}, {Mikhailov},
  {Molodzov}, {Mzhelsky}, {Mueller}, {Nandra}, {Nazarov}, {Pavlinsky},
  {Poghodin}, {Predehl}, {Robrade}, {Sazonov}, {Scheuerle}, {Shirshakov},
  {Tkachenko}, \& {Voron}}]{Sunyaev2021}
{Sunyaev}, R., {Arefiev}, V., {Babyshkin}, V., {et~al.} 2021, arXiv e-prints,
  arXiv:2104.13267

\bibitem[{{Traulsen} {et~al.}(2020){Traulsen}, {Schwope}, {Lamer}, {Ballet},
  {Carrera}, {Ceballos}, {Coriat}, {Freyberg}, {Koliopanos}, {Kurpas},
  {Michel}, {Motch}, {Page}, {Watson}, \& {Webb}}]{Traulsen2020}
{Traulsen}, I., {Schwope}, A.~D., {Lamer}, G., {et~al.} 2020, \aap, 641, A137

\bibitem[{{Trifonov} {et~al.}(2019){Trifonov}, {Rybizki}, \&
  {K{\"u}rster}}]{Trifonov2019}
{Trifonov}, T., {Rybizki}, J., \& {K{\"u}rster}, M. 2019, \aap, 622, L7

\bibitem[{{Truemper}(1982)}]{Truemper1982}
{Truemper}, J. 1982, Advances in Space Research, 2, 241

\bibitem[{{Turner} {et~al.}(2001){Turner}, {Abbey}, {Arnaud}, {Balasini},
  {Barbera}, {Belsole}, {Bennie}, {Bernard}, {Bignami}, {Boer}, {Briel},
  {Butler}, {Cara}, {Chabaud}, {Cole}, {Collura}, {Conte}, {Cros}, {Denby},
  {Dhez}, {Di Coco}, {Dowson}, {Ferrando}, {Ghizzardi}, {Gianotti}, {Goodall},
  {Gretton}, {Griffiths}, {Hainaut}, {Hochedez}, {Holland}, {Jourdain},
  {Kendziorra}, {Lagostina}, {Laine}, {La Palombara}, {Lortholary}, {Lumb},
  {Marty}, {Molendi}, {Pigot}, {Poindron}, {Pounds}, {Reeves}, {Reppin},
  {Rothenflug}, {Salvetat}, {Sauvageot}, {Schmitt}, {Sembay}, {Short},
  {Spragg}, {Stephen}, {Str{\"u}der}, {Tiengo}, {Trifoglio}, {Tr{\"u}mper},
  {Vercellone}, {Vigroux}, {Villa}, {Ward}, {Whitehead}, \&
  {Zonca}}]{Turner2001}
{Turner}, M.~J.~L., {Abbey}, A., {Arnaud}, M., {et~al.} 2001, \aap, 365, L27

\bibitem[{{Vanderburg} {et~al.}(2019){Vanderburg}, {Huang}, {Rodriguez},
  {Becker}, {Ricker}, {Vanderspek}, {Latham}, {Seager}, {Winn}, {Jenkins},
  {Addison}, {Bieryla}, {Brice{\~n}o}, {Bowler}, {Brown}, {Burke}, {Burt},
  {Caldwell}, {Clark}, {Crossfield}, {Dittmann}, {Dynes}, {Fulton}, {Guerrero},
  {Harbeck}, {Horner}, {Kane}, {Kielkopf}, {Kraus}, {Kreidberg}, {Law}, {Mann},
  {Mengel}, {Morton}, {Okumura}, {Pearce}, {Plavchan}, {Quinn}, {Rabus},
  {Rose}, {Rowden}, {Shporer}, {Siverd}, {Smith}, {Stassun}, {Tinney},
  {Wittenmyer}, {Wright}, {Zhang}, {Zhou}, \& {Ziegler}}]{Vanderburg2019}
{Vanderburg}, A., {Huang}, C.~X., {Rodriguez}, J.~E., {et~al.} 2019, \apjl,
  881, L19

\bibitem[{{Vidal-Madjar} {et~al.}(2003){Vidal-Madjar}, {Lecavelier des Etangs},
  {D{\'e}sert}, {Ballester}, {Ferlet}, {H{\'e}brard}, \&
  {Mayor}}]{Vidal-Madjar2003}
{Vidal-Madjar}, A., {Lecavelier des Etangs}, A., {D{\'e}sert}, J.~M., {et~al.}
  2003, \nat, 422, 143

\bibitem[{{Villarreal D'Angelo} {et~al.}(2021){Villarreal D'Angelo}, {Vidotto},
  {Esquivel}, {Hazra}, \& {Youngblood}}]{Villarreal2021}
{Villarreal D'Angelo}, C., {Vidotto}, A.~A., {Esquivel}, A., {Hazra}, G., \&
  {Youngblood}, A. 2021, \mnras, 501, 4383

\bibitem[{{Voges} {et~al.}(1999){Voges}, {Aschenbach}, {Boller},
  {Br{\"a}uninger}, {Briel}, {Burkert}, {Dennerl}, {Englhauser}, {Gruber},
  {Haberl}, {Hartner}, {Hasinger}, {K{\"u}rster}, {Pfeffermann}, {Pietsch},
  {Predehl}, {Rosso}, {Schmitt}, {Tr{\"u}mper}, \& {Zimmermann}}]{Voges1999}
{Voges}, W., {Aschenbach}, B., {Boller}, T., {et~al.} 1999, \aap, 349, 389

\bibitem[{{Voges} {et~al.}(2000){Voges}, {Aschenbach}, {Boller}, {Brauninger},
  {Briel}, {Burkert}, {Dennerl}, {Englhauser}, {Gruber}, {Haberl}, {Hartner},
  {Hasinger}, {Pfeffermann}, {Pietsch}, {Predehl}, {Schmitt}, {Trumper}, \&
  {Zimmermann}}]{Voges2000}
{Voges}, W., {Aschenbach}, B., {Boller}, T., {et~al.} 2000, \iaucirc, 7432, 3

\bibitem[{{Watson} {et~al.}(1981){Watson}, {Donahue}, \& {Walker}}]{Watson1981}
{Watson}, A.~J., {Donahue}, T.~M., \& {Walker}, J.~C.~G. 1981, \icarus, 48, 150

\bibitem[{{Watson} {et~al.}(2009){Watson}, {Schr{\"o}der}, {Fyfe}, {Page},
  {Lamer}, {Mateos}, {Pye}, {Sakano}, {Rosen}, {Ballet}, {Barcons}, {Barret},
  {Boller}, {Brunner}, {Brusa}, {Caccianiga}, {Carrera}, {Ceballos}, {Della
  Ceca}, {Denby}, {Denkinson}, {Dupuy}, {Farrell}, {Fraschetti}, {Freyberg},
  {Guillout}, {Hambaryan}, {Maccacaro}, {Mathiesen}, {McMahon}, {Michel},
  {Motch}, {Osborne}, {Page}, {Pakull}, {Pietsch}, {Saxton}, {Schwope},
  {Severgnini}, {Simpson}, {Sironi}, {Stewart}, {Stewart}, {Stobbart}, {Tedds},
  {Warwick}, {Webb}, {West}, {Worrall}, \& {Yuan}}]{Watson2009}
{Watson}, M.~G., {Schr{\"o}der}, A.~C., {Fyfe}, D., {et~al.} 2009, \aap, 493,
  339

\bibitem[{{Weisskopf} {et~al.}(2002){Weisskopf}, {Brinkman}, {Canizares},
  {Garmire}, {Murray}, \& {Van Speybroeck}}]{Weisskopf2002}
{Weisskopf}, M.~C., {Brinkman}, B., {Canizares}, C., {et~al.} 2002, \pasp, 114,
  1

\bibitem[{{Wood} {et~al.}(2018){Wood}, {Laming}, {Warren}, \&
  {Poppenhaeger}}]{Wood2018}
{Wood}, B.~E., {Laming}, J.~M., {Warren}, H.~P., \& {Poppenhaeger}, K. 2018,
  \apj, 862, 66

\bibitem[{{Yelle}(2004)}]{Yelle2004}
{Yelle}, R.~V. 2004, \icarus, 170, 167

\bibitem[{{Zhou} {et~al.}(2021){Zhou}, {Quinn}, {Irwin}, {Huang}, {Collins},
  {Bouma}, {Khan}, {Landrigan}, {Vanderburg}, {Rodriguez}, {Latham}, {Torres},
  {Douglas}, {Bieryla}, {Esquerdo}, {Berlind}, {Calkins}, {Buchhave},
  {Charbonneau}, {Collins}, {Kielkopf}, {Jensen}, {Tan}, {Hart}, {Carter},
  {Stockdale}, {Ziegler}, {Law}, {Mann}, {Howell}, {Matson}, {Scott}, {Furlan},
  {White}, {Hellier}, {Anderson}, {West}, {Ricker}, {Vanderspek}, {Seager},
  {Jenkins}, {Winn}, {Mireles}, {Rowden}, {Yahalomi}, {Wohler}, {Brasseur},
  {Daylan}, \& {Col{\'o}n}}]{Zhou2021}
{Zhou}, G., {Quinn}, S.~N., {Irwin}, J., {et~al.} 2021, \aj, 161, 2

\end{thebibliography}




\appendix

\section{X-ray fluxes of host stars with nearby stellar companions}

Information on flux corrections applied for X-ray data from exo\-pla\-net host stars with nearby stellar companions is given below. We use information on stellar companions from \citet{Mugrauer2019}. We checked all host stars with a known stellar companion within 50$^{\prime\prime}$ for the X-ray instruments they were detected with, and considered a multi-star system to be likely blended in eROSITA/ROSAT/\textit{XMM-Newton}/\textit{Chandra} if the stellar separation is below 8/30/8/1$^{\prime\prime}$, respectively. In those cases we divided the final listed X-ray flux, which we use for the calculations of the planetary irradiation and mass loss, by the number of blended stars (typically two). If a star is close to or below the blending separation, but X-ray data from a telescope with higher spatial resolution was available, we used the high-resolution data as the final listed flux.

\textit{18 Del}: This star has a known stellar companion at a separation of about 29.2$^{\prime\prime}$, and the only available X-ray detection stems from ROSAT, which does not resolve the two stars. We have therefore assigned 50\% of the detected ROSAT flux at the position of this binary system to the exoplanet host star.

\textit{2MASS J01033563-5515561 A}: This star has a known stellar companion at a separation of about 2$^{\prime\prime}$, and the existing XMMNewton and eRASS detections do not resolve the system. We therefore assigned 50\% of the detected eRASS flux at the position of this binary system to the exoplanet host star.

\textit{CoRoT-2}: This star has a known stellar companion at a separation of about 4.1$^{\prime\prime}$. The existing \textit{Chandra} observations resolves the system, the \textit{XMM-Newton} observations does not. However, the \textit{Chandra} data showed that the companion star is very X-ray faint and does not significantly contribute to the total X-ray flux of the system\citep{Schroeter2011}, so that no adjustment was necessary.

\textit{DS Tuc A}: This star has a known stellar companion at a separation of about 5$^{\prime\prime}$ \citep{Newton2019}. The current eRASS catalog does not resolve the two stars, and we have therefore assigned 50\% of the detected eRASS flux at the position of this binary system to the exoplanet host star.

\textit{GJ 338 B}: This star has a known, optically brighter stellar companion at a separation of 17$^{\prime\prime}$. The only existing X-ray detection stems from ROSAT, which does not resolve this wide binary. We have therefore assigned 50\% of the detected \textit{XMM-Newton} flux at the position of this system to the exoplanet host star.

\textit{HAT-P-16}: This is a hierarchical triple system, with a close stellar companion known at a separation of 0.4$^{\prime\prime}$ from the planet host star, and another wide companion at a separation of 23.3$^{\prime\prime}$. The close AB system is not resolved in the existing \textit{XMM-Newton} detection, but the wider C component is not blended. We have therefore assigned 50\% of the detected \textit{XMM-Newton} flux at the position of this system to the exoplanet host star.

\textit{HD 103774}: This star has a known stellar companion at a separation of about 6.2$^{\prime\prime}$, and X-ray detections are present from eRASS1 and eRASS2, which do not resolve the two stars. We have therefore assigned 50\% of the detected eROSITA flux at the position of this binary system to the exoplanet host star.

\textit{HD 142}: This star has a known stellar companion at a separation of about 3.9$^{\prime\prime}$, and X-ray detections are present from eRASS1 and eRASS2, which do not resolve the two stars. We have therefore assigned 50\% of the detected eROSITA flux at the position of this binary system to the exoplanet host star.

\textit{HD 162004}: This star, also known as $\psi$ 1 Dra B, has a known stellar companion at a separation of about 30$^{\prime\prime}$, and an X-ray detections is only present from ROSAT, which does not fully resolve the two stars. We have therefore assigned 50\% of the detected ROSAT flux at the position of this binary system to the exoplanet host star.

\textit{HD 189733}: This star has a known stellar companion at a separation of about 11.4$^{\prime\prime}$. The ROSAT data do not resolve the system, but the \textit{Chandra} and \textit{XMM-Newton} observations do. Furthermore, it is known from an analysis of the \textit{Chandra} data \citep{Poppenhaeger2013} that the stellar companion is much less X-ray bright than the planet host star, therefore no adjustment was necessary.

\textit{HD 195019}: This star has a known stellar companion at a separation of about 3.4$^{\prime\prime}$, and X-ray detections are present from eRASS2, which does not resolve the two stars. We have therefore assigned 50\% of the detected eROSITA flux at the position of this binary system to the exoplanet host star.

\textit{HD 19994}: This is a triple system where the planet host star is orbited by a close binary system (components B and C) at a separation of about 2.2$^{\prime\prime}$. X-ray detections are present from eRASS1 and eRASS2, which do not resolve the three stars. We have therefore assigned 1/3 of the detected eROSITA flux at the position of this system to the exoplanet host star.

\textit{HD 212301}: This star has a known stellar companion at a separation of about 4.4$^{\prime\prime}$, and an X-ray detection is present from eRASS1, which does not resolve the two stars. We have therefore assigned 50\% of the detected eROSITA flux at the position of this binary system to the exoplanet host star.

\textit{HD 65216}: This is a triple system where the planet host star is orbited by a close binary system (components B and C) at a separation of about 7.2$^{\prime\prime}$. X-ray detections are present from eRASS1 and eRASS2, which do not resolve the three stars. We have therefore assigned 1/3 of the detected eROSITA flux at the position of this system to the exoplanet host star.

\textit{HIP 65 A}: This star has a known stellar companion at a separation of about 4$^{\prime\prime}$, and the existing X-ray detections from eRASS do not resolve the stars. We therefore assigned 50\% of the detected eROSITA flux at the position of this system to the exoplanet host star.

\textit{HR 858}: HR~858 is a late-F type star that was reported to have a co-moving stellar companion of spectral type M at a separation of 8.4$^{\prime\prime}$ \citep{Vanderburg2019}. An X-ray source found in the eRASS datasets is located at the position of the secondary star and not the planet-hosting primary, which is why we attribute the detected X-ray flux to the secondary alone and do not report an X-ray detection for the planet host star based on the available data.

\textit{Kepler-1651}: This star has a known stellar companion at a separation of about 4.1$^{\prime\prime}$, and an X-ray detection is present from ROSAT, which does not resolve the two stars. We have therefore assigned 50\% of the detected ROSAT flux at the position of this binary system to the exoplanet host star.

\textit{LTT 1445 A}: This star has a known stellar companion at a separation of about 6.7$^{\prime\prime}$. The eRASS data detects flux from the position of the secondary only, and we attribute the detected X-ray flux to the companion star alone.

\textit{$\tau$ Boo}: This star has a close stellar companion at a separation of about 2$^{\prime\prime}$. X-ray detections exist with several X-ray missions, including \textit{Chandra}. The existing \textit{Chandra} observation is able to resolve the system, and shows that the secondary is much fainter than the planet-hosting primary \citep{Wood2018}, so that no adjustment was necessary.

\textit{WASP-140}: This star has a known stellar companion at a separation of about 7.2$^{\prime\prime}$, and an X-ray detection is present from eRASS2, which does not resolve the two stars. We have therefore assigned 50\% of the detected eROSITA flux at the position of this binary system to the exoplanet host star.

\textit{WASP-8}: This star has a known stellar companion at a separation of about 4.5$^{\prime\prime}$, and an X-ray detection is present from eRASS1 and \textit{Chandra}. {\textit{Chandra} observations have shown that the secondary is X-ray faint compared to the primary \citep{Salz2015massloss}, and we therefore attribute the detected eRASS1 flux to the primary in our further calculations.

\textit{omi UMa}: This star has a known stellar companion at a separation of about 6.8$^{\prime\prime}$, and an X-ray detection is present from ROSAT, which does not resolve the two stars. We have therefore assigned 50\% of the detected ROSAT flux at the position of this binary system to the exoplanet host star.




\section{Table of exoplanet irradiation fluxes and estimated mass loss rates}
We show an excerpt from the electronic data table with the most interesting columns for exoplanetary considerations in Table~\ref{tab:datatable}.

\pagebreak
\clearpage

\begingroup
\let\clearpage\relax 
\onecolumn 

\begin{sidewaystable}[t]
\caption{Excerpt from the available electronic data table; the full table has additional columns and a total of 343 rows. Radius superscript $A$ denotes planets which had no measured radius values whose radii were estimated according to the mass-radius relationship by \cite{Chen2017}. The same method was used to estimate planet masses with no measured value which are denoted with superscript $B$ in the electronic table. Superscript $C$ for the X-ray flux denotes that the host star's adopted X-ray flux and all entries derived from it were adjusted for unresolved bound stellar companions. The provenance of the used X-ray flux is given as E/R/X/C for eROSITA, ROSAT, \textit{XMM-Newton}, and \textit{Chandra}, respectively.}
\label{tab:datatable}
\begin{center}
\small
\renewcommand{\arraystretch}{1.5}
\begin{tabular}{cccccccccccc}
\hline\hline
Planet & Distance & Semi-major Axis &M$_{pl}$ & R$_{pl}$  & X-ray & $F_X$ &  $L_X$ & $L_{XUV}$   & $F_{\mathrm{X,\,pl}}$  & $F_{\mathrm{XUV,\,pl}}$  & Mass-Loss \\
& [pc] & [AU] & [$M_\oplus$] & [$R_\oplus$] & provenance & [erg\,s$^{-1}$cm$^{-2}$] & [\,erg\,s$^{-1}$] & [\,erg\,s$^{-1}$]  & [erg\,s$^{-1}$cm$^{-2}$] & [erg\,s$^{-1}$cm$^{-2}$] & [g\,s$^{-1}$] \\
\hline

11 Com b & 93.2 & 1.29 & 6165.6 & 12.4$^A$ & E & 2.8e-14 & 2.9e+28 & 2.5e+29 & 6.3e+00 & 5.3e+01 & 6.1e+06 \\
14 Her b & 17.9 & 2.93 & 1481.1 & 13.1$^A$ & X & 2.4e-14 & 9.4e+26 & 1.2e+28 & 3.9e-02 & 5.1e-01 & 2.9e+05 \\
18 Del b & 76.2 & 2.60 & 3273.5 & 12.7$^A$ & R & 2.5e-13$^C$ & 1.7e+29 & 1.2e+30 & 9.2e+00 & 6.3e+01 & 1.5e+07 \\
1RXS J160929.1-210524 b & 139.1 & 330.00 & 3000.0 & 12.8$^A$ & E & 2.3e-13 & 5.4e+29 & 3.3e+30 & 1.8e-03 & 1.1e-02 & 2.7e+03 \\
2MASS J01033563-5515561 AB b &  & 84.00 &  &  & E & 7.8e-14$^C$ &  &  &  &  &  \\
2MASS J01225093-2439505 b & 33.8 & 52.00 & 7786.5 & 12.3$^A$ & E & 4.8e-13 & 6.5e+28 & 5.1e+29 & 8.6e-03 & 6.6e-02 & 5.9e+03 \\
2MASS J02192210-3925225 b &  & 156.00 &  &  & E & 5.3e-14 &  &  &  &  &  \\
30 Ari B b & 44.7 & 0.99 & 4392.4 & 12.6$^A$ & R & 2.9e-12 & 6.9e+29 & 4.0e+30 & 2.5e+02 & 1.5e+03 & 2.5e+08 \\
51 Eri b & 29.8 & 12.00 &  &  & E & 2.4e-13 & 2.5e+28 & 2.2e+29 & 6.2e-02 & 5.4e-01 &  \\
55 Cnc b & 12.6 & 0.11 & 255.4 & 14.1$^A$ & E & 4.6e-14 & 8.7e+26 & 1.1e+28 & 2.4e+01 & 3.2e+02 & 1.3e+09 \\
55 Cnc c & 12.6 & 0.24 & 51.2 & 8.4$^A$ & E & 4.6e-14 & 8.7e+26 & 1.1e+28 & 5.5e+00 & 7.2e+01 & 3.1e+08 \\
55 Cnc d & 12.6 & 5.96 & 991.6 & 13.3$^A$ & E & 4.6e-14 & 8.7e+26 & 1.1e+28 & 8.7e-03 & 1.1e-01 & 1.0e+05 \\
55 Cnc e & 12.6 & 0.02 & 8.0 & 1.9 & E & 4.6e-14 & 8.7e+26 & 1.1e+28 & 1.3e+03 & 1.7e+04 & 5.2e+09 \\
55 Cnc f & 12.6 & 0.77 & 47.8 & 8.0$^A$ & E & 4.6e-14 & 8.7e+26 & 1.1e+28 & 5.2e-01 & 6.9e+00 & 2.7e+07 \\
61 Vir b & 8.5 & 0.05 & 5.1 & 2.1$^A$ & E & 7.2e-14 & 6.2e+26 & 8.6e+27 & 8.8e+01 & 1.2e+03 & 8.6e+08 \\
61 Vir c & 8.5 & 0.22 & 18.2 & 4.5$^A$ & E & 7.2e-14 & 6.2e+26 & 8.6e+27 & 4.7e+00 & 6.4e+01 & 1.2e+08 \\
61 Vir d & 8.5 & 0.48 & 22.9 & 5.2$^A$ & E & 7.2e-14 & 6.2e+26 & 8.6e+27 & 9.8e-01 & 1.3e+01 & 3.1e+07 \\
AB Pic b & 50.0 & 260.00 & 4290.5 & 12.6$^A$ & E & 2.4e-12 & 7.2e+29 & 4.2e+30 & 3.8e-03 & 2.2e-02 & 3.8e+03 \\
... & ... & ... & ... & ... & ... & ... & ... & ... & ... & ... \\

\hline\hline
\end{tabular}
\end{center}
\end{sidewaystable}
\endgroup

\end{document}